\documentclass[amsmath,twocolumn,superscriptaddress,showpacs,aps,prb]{revtex4}
\usepackage{bm}
\usepackage{graphics}

\def\bE{{\bf E}}
\def\bb{{\bf b}}
\def\bk{{\bf k}}
\def\bs{{\bf s}}
\def\bj{{\bf j}}
\def\br{{\bf r}}
\def\bv{{\bf v}}
\def\bM{{\bf M}}
\def\eq{{\rm eq}}
\def\bsymbol#1{\mbox{\boldmath$\displaystyle#1$}}
\def\bsig{\bsymbol{\sigma}}
\def\bPhi{\bsymbol{\Phi}}
\begin{document}
\title{Small-angle impurity scattering and the spin Hall conductivity
       in 2D systems}

\author{A. V. Shytov}
\affiliation{Physics Department, Brookhaven National Laboratory, 
             Upton, NY 11973-5000}
\author{E. G.  Mishchenko}
\affiliation{Department of Physics, University of Utah, 
             Salt Lake City, UT 84112}
\author{H.-A. Engel}
\affiliation{Lyman Laboratory, Physics Department,
             Harvard University, Cambridge, MA 02138}
\author{B. I. Halperin}
\affiliation{Lyman Laboratory, Physics Department, 
             Harvard University, Cambridge, MA 02138}
\date{\today}


\begin{abstract}
An arbitrarily small concentration of  impurities can affect the
spin Hall conductivity in a two-dimensional semiconductor system. We
develop a Boltzmann-like equation that can be used for impurity
scattering with arbitrary angular dependence, and for
arbitrary
angular dependence of the spin-orbit field  $\bb ( \bk )$ around the
Fermi surface.  For a model applicable to  a 2D hole system in GaAs,
if the impurity scattering is not isotropic, we find that the spin
Hall conductivity
depends on the derivative of $b$ with respect to the energy and on
deviations from a parabolic band structure, as well as on the
angular dependence of the scattering.  In principle, the resulting spin Hall
conductivity can be larger or smaller than the ``intrinsic value'',
and can have opposite sign.
In the limit of small angle
scattering, in a model appropriate for small hole concentrations, where the
band is
parabolic and $b \propto k^3$, 
the spin Hall conductivity has
opposite sign from the intrinsic value, and has larger magnitude.
Our
analysis assumes that the spin-orbit splitting $b$ and the transport
scattering rate $\tau^{-1}$ are both small  compared to the Fermi
energy, but the method is valid for for arbitrary value of $b \tau$.
\end{abstract}
\pacs{72.25.-b, 73.23.-b, 73.50.Bk}
\maketitle

 \section{Introduction}

An electric current passing through a semiconductor can induce
spin polarization near lateral edges of the sample, with opposite sign
at opposite edges. Such an effect, commonly
referred to as the {\it spin Hall effect}, 
had been predicted more than 30 years ago~\cite{DP71}
and was observed recently.\cite{Kato04,Wunderlich04,Sih05}
This effect results from  the coupling between 
spin and momentum of an itinerant electron.
Two specific mechanisms have been considered for the spin-Hall effect.
The ``extrinsic'' mechanism couples spin to momentum during events of
impurity scattering in the Mott skew scattering
channel~\cite{DP71,H99,Zhang00}.  As a result, scattered electrons of
different spin polarizations propagate towards opposite boundaries
of the system. As opposed to the extrinsic, the ``intrinsic'' effect
originates from a spin-split band structure~\cite{Murakami03,Sinova04}
rather than from impurity scattering.

Experiments have been performed on both electron-doped~\cite{Kato04,Sih05}
and hole-doped~\cite{Wunderlich04} GaAs-based semiconductor structures
of different dimensionalities. A spin Hall effect in three-dimensional
$n$-doped GaAs films \cite{Kato04} has been explained in terms of
extrinsic  skew and side-jump impurity scattering~\cite{Engel05,Tse05}
while the intrinsic mechanism was estimated~\cite{Bernevig04} to yield
spin accumulation smaller by an  order of magnitude than actually
observed  in the experiment. The two-dimensional electron gas on a
(110) surface of GaAs, studied in [\onlinecite{Sih05}],  was also in
the dirty limit, and the spin Hall effect observed there was
attributed  to extrinsic effects.  On the other hand, $p$-doped
two-dimensional (2D) samples used in the experiments
of Ref.~[\onlinecite{Wunderlich04}] were estimated to contain
very few impurities and it was suggested that the observations reflect
a spin-Hall effect of the intrinsic type.

While experiments directly measure  spin accumulation along the
sample edges, theoretical contributions to the field of spin-Hall
effect are mostly concerned with a different quantity which is
easier to calculate, namely, the spin current. The latter is
conventionally defined as the expectation value of the operator
$\hat{j}^{i}_{k} = C  \{ \hat s_i, \hat v_k \}$, where $\hat {\bf
s}$ and $\hat {\bf v}$ are, respectively, the operators of the
carrier spin and velocity (which is, in general, also
spin-dependent), and the normalization constant $C$ has been chosen
differently by different authors. (In the present paper,  we define
 $\hat{s}_i$ to have eigenvalues $\pm \hbar/2$, and we choose
$C=1/2$, so that the spin current is equal to $\hbar/2$ times the
difference in the particle currents for carriers  with opposite
spins.) Apart from the normalization, the definition of spin current
is largely a matter of convenience since it
{\it is not} rigorously related to spin polarization (accumulation)
by a conservation law, as opposed to {\it e.g.} the electric charge
conservation, namely, $\partial_t {s_i}+\nabla_k { j}^i_{k}\ne 0$.
Non-conservation of spin-current (due to spin precession) makes the
problem of its conversion into polarization highly non-trivial.
Moreover, since spin current is even with respect to time inversion,
it is not necessarily absent even in equilibrium \cite{Rashba03}. In
the present paper we  concentrate on the 2D electron and hole
systems. In this case the components of spin current $j^z_k$
polarized along the direction ($z$) perpendicular to the 2D plane
appear due to non-equilibrium conditions only.

The paper [\onlinecite{Sinova04}] predicted a universal value for the
intrinsic spin-Hall current in a  2D electron
gas, with spin-orbit coupling of the Rashba form,  which, using the
normalization of the present paper is given by
\begin{equation}
\label{sin}
 j_y^z=-\frac{eE_x}{8\pi},
\end{equation}
independent of the strength of spin-orbit interaction. (The
prediction of Ref. [\onlinecite{Murakami03}] for {\it three-dimensional}
hole systems, though dependent on the electron density, was also
independent  of the strength of SO
coupling). These results immediately posed a question of how
impurities would affect the spin-Hall conductivity. The derivation of formula
(\ref{sin}) neglected scattering, but also assumed a steady
state. The latter, however, is impossible in an infinite system if
momentum relaxation is absent. 
Soon a number of papers appeared
about  the role of impurity scattering, which, initially, reached differing
conclusions. Some studies encountered decreasing of the magnitude of
the effect with increasing scattering~\cite{Schliemann04,Burkov04},
while others found a universal value (\ref{sin}) even in the
presence of scattering. However, it is now generally agreed that
in the simplest model of a 2DEG, with a spin-orbit coupling that is
linear  in the wavevector $\bk$, the dc spin Hall conductivity will
vanish, even in the case of arbitrarily small impurity concentration.
Inoue et al.\cite{Inoue04} first reached this conclusion
in the clean limit, $b \tau \to \infty$, where the spin orbit
splitting  $b$ exceeds the transport scattering rate $\tau^{-1}$,
while entertaining the possibility of a finite effect for finite
values  of $b\tau$.

The present authors~\cite{Mishchenko04} developed a theory based on
a quantum kinetic equation to describe spin-polarized
transport in the two-dimensional electron systems for
{\it arbitrary} ratio of spin-orbit splitting  and scattering rate, as
long as they are both small compared to the Fermi energy ($b,\tau^{-1}
\ll E_F$). Applying this theory to the spin-Hall effect in a 2DEG
with (${\bf k}$-linear) Rashba coupling and isotropic impurity scattering, we
established the vanishing of spin Hall conductivity for arbitrary value of
the product $b\tau$.  However, the cancellation is complete only in the bulk.
Near the contacts which inject unpolarized electrons, spin-currents
are non-zero and, hence, a spin-polarization normal to plane can
accumulate  near the corners of a sample.

The models used in Refs. [\onlinecite{Inoue04}] 
and~[\onlinecite{Mishchenko04}] assumed  short-range impurity
potentials, with isotropic scattering. The result of vanishing spin
Hall conductivity in a 2DEG with Rashba coupling was subsequently
confirmed by several studies
\cite{Khaetskii04,Raimondi05,Rashba04,Dimitrova04}, and were shown
to hold for arbitrary angular dependence of the impurity scattering.
Early numerical calculations \cite{Nomura04,Murakami05} seemed to
contradict the theoretical work but later ones supported the
vanishing of spin-Hall current \cite{Nomura05}. Results for a 2DEG
with Rashba coupling can also be applied directly to a model for a
2DEG on an (001) surface with pure ${\bf k}$-linear Dresselhaus coupling, as
the two problems are related by a rotation in spin space.

As was pointed out by Dimitrova\cite{Dimitrova04}, the vanishing of
 {\it dc} spin currents $j^z_k$
in an {\it infinite} 2D system with
$\bk$-linear spin-orbit coupling has a simple explanation in terms
of the following operator argument. Consider a system with the
spin-independent disorder and the ``Rashba''-type spin-orbit
coupling \cite{Vasko79} $H_{SO}=\alpha (\sigma_x k_y-\sigma_y k_x)$,
where $\bsymbol{\sigma}$ represents the set of three Pauli spin
matrices, and $\bk$ is the canonical momentum. The equation of
motion for the operator of spin polarization yields ($l=x,y$):
$\partial_t \sigma_l=i[H_{SO},\sigma_l]/\hbar=-4\alpha m
j^z_l/\hbar^3$, where $m$ is the electron effective mass. It follows
from this identity that a nonzero expectation value of the spin
current $j^z_l$ would result in a time-dependent spin polarization
along the corresponding direction $l$, which is impossible in a
steady state. This operator argument applies also to a model with a
combination of Rashba and $\bk$-linear Dresselhaus coupling. In
agreement with this argument, analytical calculations confirm the
vanishing of weak-localization (of the first order in $1/E_F\tau)$
corrections to the spin Hall conductivity~\cite{Chalaev04}.

 It should be emphasized that this vanishing of the
spin Hall conductivity is characteristic of the 2D electron system
with SO coupling {\it linear} in momentum. Cubic Dresselhaus terms
in the spin-orbit Hamiltonian, which are $\propto \sigma_x
k_x k_y^2-\sigma_y k_y k_x^2$ for a 2D electron system on a (001)
surface,  should result in a finite spin Hall conductivity~\cite{Malshukov04}.

 For a 2D {\it hole} system in an asymmetric confining well, the Rashba effect
gives rise to a spin orbit field $\bb(\bk)$ whose direction winds three times
around a circle in the x-y plane as $\bk$ moves once around the Fermi circle.
It turns out, in this case, that for short-range impurities there is
no cancellation in the clean limit, and one then obtains the full
value  of the intrinsic spin Hall effect, at least in a simplified
model  which neglects cubic anisotropy.\cite{Murakami04}
Curiously, in contrast to the 2DEG, it is the vertex corrections
that vanish in this case, analogous to the absence of the vertex
corrections  to the electrical conductivity for short-range
impurities~\cite{AGD}.

Vertex corrections are also absent in an $n$-doped
3D system with Dresselhaus interaction,
$H_{SO}=\lambda \sigma_x k_x(k_y^2-k_z^2)+ {\rm cycl.~ permut.}$,
and short-range impurity scattering,  in this case due to the cubic
crystal symmetry. In the absence of vertex corrections, the
impurities lead to a smooth relaxational suppression of the
spin-Hall conductivity due to the broadening of the spectral
function of carriers.
In principle, one should obtain the full intrinsic value of the
spin-Hall conductivity in the clean limit,  but 3D systems are always
in the dirty limit, where the intrinsic spin Hall conductivity is
reduced by a factor $\sim (b \tau)^2$.  As mentioned above, the spin
Hall  effect observed in 3D $n$-type GaAs is most probably due to
extrinsic scattering effects, omitted from the model discussed
here.\cite{Kato04,Engel05,Tse05}

It is natural to ask whether the absence of vertex corrections for the
2D hole system is special to the assumption of short range impurities,
or whether it applies equally well to a model with arbitrary angular
dependence of the impurity scattering. In the present paper, we
develop equations to calculate the spin Hall conductivity for arbitrary
form of the impurity potential, and for arbitrary form of the
spin-orbit field  $\bb(\bk)$ in a 2D system.  Solution of these
equations is simple in models which have an overall circular symmetry,
and we obtain analytic results in the limit of small angle scattering,
resulting from smooth disorder. This type of disorder
is a good assumption for many 2D systems in which the impurity
centers (dopants) reside relatively far from the heterostructure
interface.

The result of our analysis is that the spin-Hall conductivity, in general,
 does depend on the form of the impurity scattering, and
 except for the case of isotropic scattering, depends also on the
energy-dependence of the spin-orbit coupling and on the deviation
from parabolicity
 of the energy dispersion for the carriers.
In our circularly symmetric model, the
magnitude of the spin-orbit field $\bb(\bk)$ is independent
of the direction of $\bk$,
while its direction winds $N$ times around the unit circle in the
$x$-$y$ plane, at a uniform rate, as $\bk$ moves around the
Fermi circle. ($N$ must be an odd integer.)
The energy-dependence of $b$ and deviations of the energy dispersion
$\epsilon_k$ from a parabolic form near the Fermi energy, are then
characterized by two parameters $\widetilde{N}$ and $\zeta$, defined in
Eq. (\ref{B235}) below, and for a given form of the impurity scattering,
the spin Hall conductivity is found to
depend on $N$, $\widetilde{N}$ and $\zeta$.

In Section II, below, we define precisely the models we are
considering and the relevant parameters.  
In Section III, the transport quantities are expressed in terms of the Wigner
distribution function, for which Boltzmann-like equations of motion are presented
in section IV. In Section V,  these equations  are solved in the case of our circularly symmetric model.
In Section VI, we obtain explicit formulas, Eqs. (\ref{C8}) and (\ref{C9}), for the two limiting cases of isotropic scattering and small angle scattering by the impurities. The relation of our results to previous theoretical results and to recent experiments is discussed in Section VII. Some details of the derivations are presented in Appendix A, and an alternate derivation of the kinetic equations of motion, using Green's function methods, is presented in Appendix B.

\section{Model}

We consider a two-dimensional system of non-interacting electrons or
holes, described by a one-body Hamiltonian of the form:
\begin{equation}
\label{B1} H=\epsilon_k-\frac{1}{2} \bb(\bk)\cdot \bsymbol{\sigma}
+V(\br),
\end{equation}
where $V$ is the potential due to  impurities, and $\bk$ is the
kinetic momentum operator.  In the presence of a vector potential
${\bf A}  $, we have $\bk = {\bf p} - e {\bf A} / c$, where ${\bf p}
\equiv  - i\nabla $.  Throughout this paper, $e$ is the charge of
the carriers, and we set $\hbar = 1$ except where explicitly noted.
We may include a uniform electric field $\bE$ by letting ${\bf A}$
depend on time.

By time reversal symmetry, the spin-orbit  field $\bb$ must satisfy
$\bb (\bk) = - \bb ( - \bk) $ .  Note that vectors such as $\bk$ and
$\br$, which refer to spatial coordinates, have two cartesian
components, while vectors in spin space, such as $\bb$ have three
components.

We have assumed for  simplicity that $\epsilon_k$, the energy
dispersion in the absence of spin orbit coupling, is isotropic and a
monotonic function of $k$.  The formulas derived below can be
extended in a straightforward manner to more complicated band
structures, but the formulas would then be much less transparent,
and the resulting integral equations are more difficult to solve,

As a further simplification, we  shall later specialize to models
where $\bb$ has a simple rotational symmetry; i.e., we assume
\begin{equation}
\label{B22}
b_z=0, \;\;\;b_x+ib_y=b_0(k)e^{iN\theta},
\end{equation}
where $b_0(k)$ is a complex number,  whose magnitude may vary with
the magnitude of $k$, but whose  phase is independent of $\bk$. This
will allow us to obtain analytic solutions to the equations. The
final results will, of course, depend crucially on the winding
number $N$.  We shall see that results can also be affected by the
energy dependence of $b$ and by  deviations of the energy dispersion
$\epsilon_k$ from a parabolic form near the Fermi energy.  We
characterize these dependences by constants $\widetilde{N}$ and $\zeta$,
with
\begin{equation}
\label{B235} \widetilde{N} = \frac{d \ln |b_0|}{d \ln k} , \;\;\;
1+\zeta = \frac{d\ln v}{d\ln k} ,
\end{equation}
for $k$ equal to  the Fermi momentum. In other words, we assume that near
the Fermi energy, $v\propto k^{1+\zeta}$, and
\begin{equation}
\label{B24} b_x+ib_y  \propto  k^{\widetilde{N}} e^{iN\theta}.
\end{equation}

For a 2D electron system on a (100) surface of a III-V
semiconductor, the case $N=1$ corresponds to pure Rashba coupling,
while $N=-1$ corresponds to a ${\bf k}$-linear  Dresselhaus coupling.  The
case $N=3$ arises in a circularly symmetric  model of a 2D hole
system, which ignores warping due to the tetragonal symmetry. In the
limit of small hole doping, the band structure is parabolic and
$(b_x+ib_y) \propto (k_x+ik_y)^3$; hence $\zeta=0$ and
$\widetilde{N}=N=3$.  At larger values of the doping, when the
Fermi-energy becomes comparable to the splitting between the light
and heavy hole bands, the values of $\widetilde{N}$ is reduced, and the
value of $\zeta$ will be negative. At still higher doping, both
light and heavy hole bands become occupied, and the situation is
more complicated.

The Hamiltonian (\ref{B1}) assumes that there is no direct
spin-orbit coupling associated with the impurity potential or with
applied electric field.  For example, it omits terms of the form
$\lambda \bsig \cdot (\bk \times \nabla V)$, which are generally
present in systems with spin orbit coupling.  This term leads to
skew-scattering by the impurities, giving an extrinsic contribution
to the spin Hall conductivity. Since there is no direct general
relation between the coupling constant $\lambda$ and the magnitude
of the intrinsic spin-orbit splitting $b$,  it is at least logically
consistent to consider a model with $\lambda = 0$.  Moreover, the
skew scattering cross-section is higher order in the impurity
potential $V$ than the ordinary transport scattering cross-section,
so it is normally dominated by places where the carrier enters a
region of strong potential gradients, close to the impurity.  In a
remotely doped 2D hole system one might expect that the skew
scattering contribution to the spin Hall conductivity should be
relatively small, at least if one can ignore scattering by residual
impurities close to the 2D system.

We concentrate on systems which are large compared to the mean
free path. The ``mesoscopic'' spin-Hall effect
\cite{Nikolic1,Hank04,Nikolic2} in ballistic heterostructures is,
therefore, beyond the scope of the present paper.

\section{Boltzmann formulation}
\label{boltzmann-formulation}

We  begin by defining a Wigner distribution function  for the
carriers, which is a $2 \times 2$ matrix in spin space, given by
\begin{eqnarray}
\label{B2} n_{\alpha\beta}(\bk,\br,\epsilon,t)=
\frac{1}{2\pi} \int d^2{\bf s}\;d\tau
e^{i\bk\cdot\bs-i\epsilon\tau} \nonumber\\
\times \left\langle\psi_\beta^\dagger \left(\br+\frac{\bs}{2}, t
+\frac{\tau}{2}\right) \psi_\alpha \left(\br -\frac{\bs}{2}, t -
\frac{\tau}{2} \right)\right\rangle.
\end{eqnarray}
In thermal equilibrium, if spin orbit coupling is absent  ($\bb=0$)
and impurities are neglected, we have the standard result
\begin{equation}
\label{B3}
n_{\alpha\beta}=f_0(\epsilon)
\delta(\epsilon-\epsilon_k)\delta_{\alpha\beta},
\end{equation}
where $f_0$ is the Fermi function. For $b\not=0$, there are two
different wave vectors at a given energy $\epsilon$ and a given
direction  $\theta$, corresponding to spin states parallel or
antiparallel to $\bb$.  Introducing an index $\sigma = \pm 1$ to
distinguish these two states, one finds in equilibrium,
$n_{\alpha\beta}=n_{\alpha\beta}^{\eq}(k,\epsilon)$
with
\begin{equation}
\label{B4} n_{\alpha\beta}^{\eq}= \frac{1}{2} \sum_{\sigma=\pm 1}
\left(\delta_{\alpha \beta}+ \sigma \frac{
\bb\cdot\bsymbol{\sigma}_{\alpha \beta}}{b}\right) f_0(\epsilon) \,
\delta \! \left(\epsilon-\epsilon_k+\sigma \frac{b}{2}\right).
\end{equation}
Impurity scattering broadens the $\delta$ functions in these expressions,
but does not shift their centers or alter their amplitudes,
provided that the scattering rate is small compared to the
Fermi energy $E_{\rm{F}}$.

We define the distribution as a function of energy and direction
in $\bk$-space by
\begin{equation}
\label{B5} \tilde{n}_{\alpha\beta}(\theta,\epsilon) \equiv
\int_0^{\infty}  \frac{k\;dk}{(2\pi)^2} n_{\alpha\beta}(\bk,\epsilon),
\end{equation}
with $\bk=k(\cos\theta,\sin\theta)$.  (We suppress here the indices
$\br$ and $t$, and we shall omit other indices as well when the
meaning is clear.) Substituting (\ref{B4}) into (\ref{B5}), we find,
in equilibrium, to first order in $\bb / E_F$,
\begin{eqnarray}
\label{B6}
\tilde{n}_{\alpha\beta}=
\tilde{n}_{\alpha\beta}^{\eq}(\theta,\epsilon)
=\frac{k_\epsilon}{(2\pi)^2 v_\epsilon}
 f_0(\epsilon)\delta_{\alpha\beta} \nonumber\\
+f_0(\epsilon) \frac{\bsig_{\alpha \beta}}{8\pi^2}
 \cdot \frac{\partial}{\partial \epsilon}
\left(\frac{k_\epsilon \bb (k_\epsilon)}{v_\epsilon} \right),
\end{eqnarray}
where $v_\epsilon=\frac{\partial\epsilon_k}{\partial k}$, and
$k_{\epsilon}$ is the value of $k$ such that  $\epsilon_k$ is equal
to the given value of $\epsilon$.  For $\epsilon$ at the Fermi
energy, $k_\epsilon$ is the Fermi momentum in the absence of
spin-orbit coupling, and $v_\epsilon$ is the corresponding Fermi
velocity. The last term in (\ref{B6}) may be understood as arising
from the difference in the densities of states for the two spin
states $\sigma = \pm 1$, which may be written as $(2\pi)^{-2}
(k_\epsilon + \delta k) / (v_\epsilon + \delta v)$,  where $\delta k
= (b \sigma /2v_\epsilon)$ and $\delta v = (d^2 \epsilon_k / d k^2)
\delta k + (1/2) \sigma \partial b/\partial k $. The shifts in momentum and
velocity have the opposite sign for the two values of $\sigma$, and
the final term in Eq.~(\ref{B6}) is obtained by expanding to first
order in these shifts.

Finally, we introduce functions $n(\theta, \epsilon)$ and
$\bPhi(\theta,\epsilon)$, which describe the excess number and spin
densities in  momentum direction $\theta$ and energy $\epsilon$,
defined by
\begin{equation}
\label{B7} \tilde{n}_{\alpha\beta}
(\theta,\epsilon)-\tilde{n}_{\alpha\beta}^{\eq}(\theta,\epsilon)
\equiv\frac{n(\theta,\epsilon)}{2} \delta_{\alpha \beta}
+\bsymbol{\Phi}(\theta,\epsilon)\cdot
\bsymbol{\sigma}_{\alpha\beta}.
\end{equation}

If $n$ and $\bPhi$ are known, we can readily compute the particle
current and spin current,  as well as the particle density and spin
density at a spatial point $\br$. We define the spin density and
spin current by
\begin{equation}
\label{B38} s^\mu(\br)= \frac{1}{2}  \sum_{\alpha,\beta}
\langle\psi_\alpha^+(\br)\sigma_{\alpha\beta}^\mu
\psi_\beta(\br)\rangle,
\end{equation}
\begin{equation}
\label{B39}
\bj^\mu(\br)=\frac{1}{4} \sum_{\alpha, \beta}
\langle\psi_\alpha^+(\br)\{\sigma^\mu,\bv\}_{\alpha \beta}
\psi_\beta(\br)\rangle,
\end{equation}
where ${\bf v}$ is the velocity operator,
\begin{equation}
\label{B40}
\bv=\frac{\partial\epsilon_k}{\partial\bk}
-\frac{\bsymbol{\sigma}}{2}
\cdot\frac{\partial\bb(\bk)}{\partial\bk} .
\end{equation}


At any point $\br$, we  then have for densities and currents
\begin{equation}
\label{B41} \langle n\rangle-n_0=\int_0^{2\pi}
d\theta\int_{-\infty}^\infty d\epsilon\;n(\theta,\epsilon),
\end{equation}
\begin{equation}
\label{B42}
s^\mu=  \int_0^{2\pi} d\theta\int_{-\infty}^\infty d\epsilon\;
      \Phi_\mu(\theta,\epsilon),
\end{equation}
\begin{equation}
\label{B43}
\bj=\int_0^{2\pi} d\theta\int_{-\infty}^\infty d\epsilon\;
\bv_\epsilon(\theta)n(\theta,\epsilon),
\end{equation}
\begin{equation}
\label{B44}
\bj^\mu=  \int_0^{2\pi} d\theta\int_{-\infty}^\infty d\epsilon\;
[\bv_\epsilon(\theta)
\Phi_\mu(\theta,\epsilon)+ \frac{1}{2} \delta\bv_\mu(\theta)n(\theta,\epsilon)]
\ . 
\end{equation}
Here 
\begin{equation}
\label{B45} \bv_\epsilon(\theta)\equiv
v_\epsilon(\cos\theta,\sin\theta),
\end{equation}
is the ``bare'' velocity, and
\begin{equation}
\label{B46} \delta\bv_\mu(\theta)\equiv -\frac{1}{2} \frac{\partial
b_\mu(\bk)} {\partial\bk}
\end{equation}
is the spin-orbit contribution. 
In the equation for $\bj$, we have dropped terms of order $|b|\Phi_\mu$.
This is justified because we assume $|b|/E_{\rm{F}} \ll 1$,
and in the situations we will consider,  values of $|\bPhi|$ are smaller
than the characteristic values of $n(\theta,\epsilon)$
by a factor of $|b|/E_{\rm{F}} $ or more.

\section{Equations of motion}

We will be interested in the linear response of the system
to an infinitesimal applied electric field $\bE$, which might,
in general, depend on space as well as time. In developing
the equations of motion, then, we need only consider terms
which are of first order in $\bPhi$ and $n$, or first order in $E$;
we may ignore terms of order $En$ or $E \Phi$.

In the absence of spin-orbit coupling, the linearized equation of motion
for $n(\theta, \epsilon, \br, t) $ may be written as
\begin{equation}
\label{B8} \frac{\partial n}{\partial t}= \left.\frac{\partial
n}{\partial t}\right |_{v} + \left.\frac{\partial n}{\partial
t}\right |_{E} + \left.\frac{\partial n}{\partial t}\right |_{\rm
scat}
\ ,
\end{equation}
where 
\begin{equation}
\label{B9} \left.\frac{\partial n}{\partial t}\right|_v=
-v_\epsilon\hat{k}\cdot \frac{\partial}{\partial\br}\;n,
\end{equation}
is the standard advection term, 
\begin{equation}
\label{B10} \left.\frac{\partial n}{\partial t}\right|_E=
\frac{2eEk\cos\theta}{(2\pi)^2} \left( \frac{-\partial
f_0(\epsilon)} {\partial\epsilon} \right),
\end{equation}
describes acceleration of particles by the external field,
and 
\begin{equation}
\label{B11}
\left.\frac{\partial n(\theta)}{\partial t}
\right |_{\rm scat}=
\int_0^{2\pi}
\;
d\theta'\;K(\theta-\theta')[n(\theta')-n(\theta)] ,
\end{equation}
is the collision integral.
The scattering kernel $K$ is given, in the Born approximation, by
\begin{eqnarray}
\label{B12}
K(\theta-\theta')&=&W(q)k_\epsilon/(2\pi v_\epsilon),\nonumber\\
q&\equiv&2k_\epsilon\sin(|\theta-\theta'|/2),\nonumber\\
W(q)&\equiv&\langle|V({\bf q})|^2\rangle.
\end{eqnarray}
Here $V({\bf q})$ is the Fourier component of the scattering potential
at wavevector ${\bf q}$, and we have assumed that its mean-square
value is independent of the direction in space.  In (\ref{B10}),
we have assumed that the electric field $\bE$ is oriented along the
$x$-axis. The equations assume that spatial variations are slow on the
scale of the Fermi wavelength, and they may need to be supplemented
by boundary conditions derived from microscopic equations,
in the case of a sharp interface or sample edge.

The equations given above for $\partial n/ \partial t $ remain valid
in the presence of spin-orbit  coupling, to first order in
$b/E_{\rm{F}}$.  The equations of motion for $\bPhi$, however, are
affected in an essential way by a nonzero $b$.   We find here
\begin{equation}
\label{B13} \frac{\partial\bsymbol{\Phi}} {\partial t} =
\left.\frac{\partial\bsymbol{\Phi}} {\partial t}\right |_v+
\left.\frac{\partial\bsymbol{\Phi}} {\partial t}\right |_b+
\left.\frac{\partial\bsymbol{\Phi}} {\partial t}\right |_E+
\left.\frac{\partial\bsymbol{\Phi}} {\partial t}\right |_{\rm scat},
\end{equation}
\begin{equation}
\label{B14} \left.\frac{\partial\bsymbol{\Phi}} {\partial t}\right
|_v = -v_\epsilon \hat{k}\cdot \frac{\partial}{\partial\br}
\bsymbol{\Phi},
\end{equation}
\begin{equation}
\label{B15} \left.\frac{\partial\bsymbol{\Phi}} {\partial t}\right
|_b = - \bb\times\bsymbol{\Phi} + n   
\left(\bb\times\frac{\partial\bb}{\partial k}\right) \frac{1}{4 
v_\epsilon}
\end{equation}
\begin{equation}
\label{B15e} \left.\frac{\partial\bsymbol{\Phi}} {\partial t}\right
|_E = \frac{Ee}{8\pi^2v_\epsilon} \left( -\frac{\partial
f_0}{\partial \epsilon} \right) \frac{\partial}{\partial \theta } (
\bb \sin \theta ),
\end{equation}
\begin{eqnarray}
\label{B16}
\left.\frac{\partial\bsymbol{\Phi}(\theta)}
{\partial t}\right |_{\rm scat} &=&
\int_0^{2\pi}
\; d\theta'\;
K(\theta-\theta')[\bsymbol{\Phi}
(\theta')-\bPhi(\theta)]\nonumber\\[5pt]
&+&\int_0^{2\pi}
\; d\theta'\;
[{\bf M}(\theta,\theta') n
(\theta')-{\bf M}(\theta',\theta)
n (\theta)] \ . \nonumber \\[5pt]
\end{eqnarray}
Spin-orbit contribution to the scattering kernel can be divided 
into two different parts: 
\begin{equation}
\label{B17}
\bM=\bM^d+\bM^w ,
\end{equation}
where
\begin{equation}
\label{B18}
\bM^d(\theta,\theta')=
\frac{v_\epsilon K(\theta-\theta')}
{4k_\epsilon}
\frac{\partial}{\partial\epsilon}
\left[\frac{k_\epsilon\bb(k_\epsilon,\theta)}{v_\epsilon}\right] ,
\end{equation}
originates from spin-orbit correction to the density of states, 
and 
\begin{equation}
\label{B19}
\bM^w(\theta,\theta')=
\frac{[\bb(\theta)+\bb(\theta')]}
{4k_\epsilon v_\epsilon}
\tan\left(\frac{\theta-\theta'}
{2}\right)\frac{\partial K(\theta-\theta')}
{\partial\theta}
\end{equation}
is due to the spin-dependence of the momentum transfer entering the
scattering matrix element. In the case of small angle scattering, 
Eqs.~(\ref{B18}) and~(\ref{B19}), which are correct to first order in $\bb$, 
require not only that $b / E_F << 1$, but also $b < v_F \bar{q}$, 
where $\bar{q}$ is the typical scattering momentum.

A more detailed  derivation of these equations will be given in
Appendices A and B. We give here just a brief indication of the
origin of the various terms. We then
shall explore some consequences of these
equations in simple cases, which will shed additional light on  their meaning
and will provide some non-trivial consistency checks.

\subsection{Origins}

The contribution given by (\ref{B15}) may be understood as arising
from precession of the spin, induced by the spin-orbit field $\bb$.
The second term on the right hand side of this equation is zero
in any case where the direction of $\bb$ is determined by
the direction of $\bk$, independent of the magnitude of $k$, and we shall
omit it in the following. The electric field term (\ref{B15e}) reflects
the action  of $\bE$ on the equilibrium distribution (\ref{B4}), as explained
in Appendix A.

The scattering kernel $\bM$  entering (\ref{B16}) gives a
contribution to $\partial \bPhi / \partial t$ proportional to~$n$, 
which is present even when $\bPhi=0$. The term
$\bM^d$, given by (\ref{B18}), arises because there are two
different Fermi radii at a given energy $\epsilon$ and a given
direction $\theta$, corresponding to spin states parallel or
antiparallel to $\bb$, and the densities of final states are
generally different for particles scattered into these two spin
states. The second term  $\bM^w$, given by (\ref{B19}) arises when
the scattering matrix element depends on $q$, because the momentum
transfer actually depends not only on $\theta - \theta'$, but also
on the initial and final Fermi radii, and hence on the component of
the spin in the directions of  $\bb (\theta)$ and $\bb(\theta')$.

\subsection{Application to simple examples}
First consider a situation where $E=0$, and $n(\theta, \epsilon)$
is independent of $\theta$. The equations of motion for $\bPhi$ then have
a time-independent solution where
\begin{equation}
\label{D1}
\bPhi = \frac{n v_\epsilon}{4 k_\epsilon} \frac {\partial}{\partial \epsilon}
\left(\frac{k_\epsilon {\bf b}}{v_\epsilon} \right).
\end{equation}
A change in electron density independent of $\theta$ is just what one
would find if one makes  a small change in the chemical potential,
by an amount  $\delta \mu$ such that
\begin{equation}
\label{D2}
n = -\delta \mu \frac{v_\epsilon k_\epsilon}{2 \pi^2}
     \left( \frac{\partial f_0}{\partial \epsilon} \right) .
\end{equation}
The change in the equilibrium distribution produced by a shift $\delta\mu$
can be obtained from  (\ref{B6}) by replacing
$f_0(\epsilon)$ by $f_0(\epsilon - \delta \mu )$.
We see that this would  change the polarization by an amount precisely
equivalent to (\ref{D1}), as is required.

As another example, we may consider a situation where there is no
impurity scattering as well as no electric field. We may then construct
spatially-varying time-independent solutions of the equations of motion
of the form
\begin{equation}
\label{B3a}
\bPhi(\br, \theta, \epsilon) = a \hat{b} + c \left[ \hat{g}
    \cos (\delta \bk \cdot \br)
+   (\hat{k} \times \hat{g}) \sin(\delta \bk \cdot \br) \right] ,
\end{equation}
where $\hat{b}$ is a unit vector parallel to ${\bf b}(k_\epsilon, \theta)$,
$\hat{g}$ is a unit vector perpendicular to $\hat{b}$,
$a$ and $c$ are arbitrary constants, which may depend on $\theta$
and $\epsilon$, and
\begin{equation}
\delta \bk = \frac {|b|}{2 v_\epsilon} (\cos \theta, \sin \theta) .
\end{equation}
This steady state corresponds to a situation where we have populated
eigenstates of the Hamiltonian with energy $\epsilon$ and wavevector-direction
$\theta$, whose spin  at $\br = 0$ is polarized in direction
$a\hat{b} + c \hat{g}$. The spin polarization precesses with
a spatial-frequency $\delta \bk$ as one moves away from the origin.

Finally, we may consider a situation where impurities exist only
in a small region about the origin,  whose radius is small compared
to $1/\delta k$ but large on the scale of the Fermi wavelength.
We  can then construct a scattering wave solution of the Hamiltonian,
for which there is an incident plane wave of definite energy $\epsilon$
and wavevector $\bk$, with polarization parallel or antiparallel to $\bb(\bk)$.
  The scattered wave radiating from the origin will have spin parallel
to the spin of the incident wave for $\br$ close to the origin,
but will have a spin direction that depends on $\br$ away from the origin.
If one moves out along a constant direction $\hat{r}$, the spin direction
will precess about the direction $\bb(\bk')$, where $\bk'$ is a vector
of magnitude $k_\epsilon$ and direction $\hat{r}$.  The polarization
$\bPhi ( \br, \theta, \epsilon)$ produced by such a scattering wave
solution will give a time-independent solution of the equations motion
derived above, at least within the Born approximation.

\section{Case of circular symmetry}

We now specialize to the case where $\bb (\bk)$ has the simple
dependence on $\theta$ given by (\ref{B22}).  We wish to find the
linear response to a uniform electric field, in an infinite
homogeneous system.

When the angular dependence of $\bb$ is given by (\ref{B22}), the
equations can be solved by Fourier transform in $\theta$, i.e., by
expanding ${\bf \Phi}(\theta)$ in circular harmonics:
\begin{eqnarray}
\label{B26}
\Phi_z&=&\sum_{m=-\infty}^\infty
\Phi_m^z e^{im\theta},\nonumber\\[5pt]
\Phi^+&\equiv&\Phi_x+i\Phi_y=\sum_m\Phi_m e^{im\theta},\nonumber\\[5pt]
K(\theta-\theta')&=&\sum_m K_me^{im(\theta-\theta')}.
\end{eqnarray}
Since $\Phi_z$ is real, we have
$\Phi_m^z=\left(\Phi_{-m}^z\right)^*$. Also, since $K$ is real and
an even function of $\theta-\theta'$,  we see that $K_m$ is real and
$K_m=K_{-m}$. Equations (\ref{B15}) and (\ref{B15e}) can then be
written as
\begin{eqnarray}
\label{B27}
\left.\frac{d\Phi_m^z}
{dt}\right |_b&=&
\frac{-b_0^*\Phi_{m+N} + b_0\Phi_{N-m}^*}{2i},\nonumber\\[5pt]
\left.\frac{d\Phi_m}
{dt}\right |_b&=&
ib_0\Phi_{m-N}^z,\nonumber\\[5pt]
\left.\frac{d\Phi_m^z}
{dt}\right |_E&=&
0,\nonumber\\[5pt]
\left.\frac{d\Phi_m} {dt}\right |_E&=&- \frac{Ee\eta_0m}{2} \left[
\delta_{m,(N+1)}-\delta_{m,(N-1)}\right] \frac{\partial
f_0}{\partial\epsilon},~~
\end{eqnarray}
\begin{equation}
\label{B30}
\eta_0 \equiv \frac{b_0}{8\pi^2v_\epsilon} .
\end{equation}

The scattering contribution depends on $n(\theta, \epsilon)$.
For a dc electric field, one obtains the standard result,
\begin{equation}
\label{B20}
n(\theta,\epsilon)=E \tau\frac{ek_\epsilon}{2\pi^2}
\left(-\frac{\partial f_0}{\partial\epsilon}\right)\cos\theta ,
\end{equation}
where $\tau$ is the transport scattering lifetime, given by
\begin{equation}
\label{B21}
\frac{1}{\tau}=2\int_{-\pi}^{\pi}\;d\theta\;K(\theta)\sin^2
\left[\frac{\theta}{2}\right].
\end{equation}
We then find
\begin{equation}
\label{B28} \left.\frac{d\Phi_m^z} {dt}\right |_{\rm scat}=
-2\pi\Phi_m^z(K_0-K_m),
\end{equation}
\begin{eqnarray}
\label{B29} &&\left.\frac{d\Phi_m}
{dt}\right |_{\rm scat}= -2\pi\Phi_m(K_0-K_m)\nonumber\\[5pt]
&& + \frac{eE}{2} \left[\gamma_{(+)}\delta_{m,(N+1)}
+\gamma_{(-)}\delta_{m,(N-1)}\right] \left(-\frac{\partial
f_0}{\partial\epsilon}\right),~~~
\end{eqnarray}
where
\begin{equation}
\label{B31}
\gamma_{(\pm)}=\gamma_0^d+\gamma_0^w\pm\lambda^w ,
\end{equation}
\begin{equation}
\label{B32}
\gamma_0^d=2 \pi \eta_0 \tau (K_1-K_N) (\widetilde{N} - \zeta ) ,
\end{equation}
\begin{equation}
\label{B33}
\gamma_0^w=2\pi\eta_0\tau
\left[\frac{\widetilde{K}_{N+1}+\widetilde{K}_{N-1}}{2}
-\widetilde{K}_N+\widetilde{K}_1-\widetilde{K}_0
\right] ,
\end{equation}
\begin{equation}
\label{B34}
\lambda^w=2\pi\eta_0\tau
\left[\frac{\widetilde{K}_{N+1}-\widetilde{K}_{N-1}}{2}
\right] ,
\end{equation}
with
\begin{eqnarray}
\label{B35}
\widetilde{K}(\theta-\theta')&\equiv&
\tan\left(\frac{\theta-\theta'}{2}\right)
\frac{\partial K(\theta-\theta')}{\partial\theta}
\nonumber\\[5pt]
&\equiv&\sum_m\widetilde{K}_m e^{i(\theta-\theta')}.
\end{eqnarray}
\vskip15pt

It is seen by inspection that the solution of these equations 
has 
all  components zero except for $\Phi_{N\pm1}$ and $\Phi^z_1 =
(\Phi^z_{-1})^* $, which are therefore determined by a set of three
coupled linear equations.  For a dc applied field, where  $d\Phi/dt
= d\Phi^z/dt =0$, these equations may be written in the form
\begin{equation}
\label{C1} 2i \Phi^z_1 = (- b_0^* \Phi_{N+1} + b_0 \Phi^*_{N-1}) \tau,
\end{equation}
\begin{equation}
\label{C2} \tau^{-1}  k_{N+1} \Phi_{N+1} - ib_0 \Phi^z_1 =
\frac{Ee}{2} \left( \frac {- \partial f_0}{\partial \epsilon }
\right) \alpha_{(+)},
\end{equation}
\begin{equation}
\label{C3}
\tau^{-1}  k_{N-1} \Phi^*_{N-1} + ib^*_0 \Phi^z_1 =
-\frac{Ee}{2}
 \left( \frac {- \partial f_0}{\partial \epsilon } \right)
 \alpha^*_{(-)} ,
\end{equation}
where
\begin{eqnarray}
\label{C4} k_m \equiv 2 \pi (K_0 - K_m) \tau , \;\;\;
\alpha_{(+)} \equiv \left[+ \gamma_{(+ )} + \eta_0 (N+1) \right],
\nonumber\\ \alpha^*_{(-)} \equiv
 \left[ -\gamma_{(-)}^* + \eta_0^{*} (N-1) \right].~
\end{eqnarray}
Note that $k_m$ are dimensionless  constants that depend on the
shape of the angular dependence of the impurity scattering, but not
on its overall strength.

Solution of these equations gives
\begin{equation}
\label{C5}
2i \Phi^z_1 = E e  \left( \frac{\partial f_0}{\partial \epsilon} \right)
\left(
  \frac{ k_{N+1} b_0 \alpha^{*}_{(-)} + k_{N-1} b_0^{*} \alpha_{(+)}}{ 2  D}
\right) ,
\end{equation}
\begin{equation}
\label{C6}
D \equiv (\tau^{-2}/2)
  \left[  |b_0|^2 \tau^2 (k_{N+1} + k_{N-1}) + 2  k_{N+1} k_{N-1}  \right] .
\end{equation}
Note that the right-hand side of (\ref{C5}) is real, so that $\Phi^z_1$
is pure imaginary.

Integrating (\ref{C5})  over the energy $\epsilon$, we find
the spin currents $j^z_x=0$ and
$j^z_y \equiv   - \sigma_{\rm{sH}} E$, with a  spin Hall conductivity given by

\begin{equation}
\label{C7} \sigma_{\rm{sH}} =  \pi e v_F
\left( \frac{ k_{N+1} b_0 \alpha_{(-)}^*
       + k_{N-1} b_0^* \alpha_{(+)} }{ 2  D}
\right) .
\end{equation}

In the clean limit, $|b| \tau \gg 1$, one can replace the denominator 
$D$ in Eq. ~(\ref{C7}) by $ |b_0|^2  (k_{N+1} + k_{N-1}) /2$.
Since $\alpha_{(\pm)} \propto |b_0|$, we see that $\sigma_{\rm{sH}}$
is independent of $\tau$ and independent of the magnitude of $|b|$
in this limit. By contrast, in the dirty limit, when $|b| \tau \ll 1$,
we see that $\sigma_{{\rm sH}} \propto |b \tau|^2 $

Note that spin-Hall conductivity cancels at $N = 1$
for an arbitrary scattering kernel $K(\theta - \theta')$. Indeed, according
to Eq.~(\ref{B32}), $\gamma_0^d$
is identically zero for $N = 1$, while $\gamma_0^w$ and $\lambda^w$
given by Eqs.~(\ref{B32}) and~(\ref{B33}) are equal.
Thus, both $\gamma^{(-)}$, and $\alpha_{(-)}$ are zero, therefore the 
first term in the Eq.~(\ref{C7}) vanish. 
Also, $k_{0} = 0$ by definition (see Eq.~(53)), 
and the spin-Hall conductivity is identically zero. 
This result is in agreement with a general argument
given in~Ref.~[\onlinecite{Dimitrova04}].

\section{Isotropic scattering and small angle scattering}

For the case of isotropic impurity scattering, we have
$K_0 = (2 \pi \tau)^{-1}$,
and $K_m = 0$ for $m \neq 0$.
We also have $\widetilde{K} =0$, $\gamma^d_0 = 0$, and $\gamma_{(\pm)} =0$.
Then, for $|N| \neq 1$, we find in the clean limit,
\begin{equation}
\label{C8} \sigma_{\rm{sH}} = \frac{e N}{8 \pi } .
\end{equation}
This is just the ``universal intrinsic value'' of the spin Hall conductivity
for this model, as has been found previously by other authors, written
for our present  normalization.  The result will be reduced by
a factor of $(1+ |b_0|^2 \tau^2)^{-1}$ if $|b_0| \tau$ is not small.

In the limit of small angle scattering, we have
\begin{eqnarray}
\label{B36}
\tau^{-1}&=&\int_{-\pi}^\pi K(\theta)
\frac{\theta^2}{2}\;d\theta, \nonumber\\[5pt]
K_0-K_m&=&\frac{m^2}{2\pi\tau}, \nonumber\\[5pt]
\widetilde{K}(\theta)&=&\frac{\theta}{2}
\frac{\partial}{\partial\theta} K(\theta)
\ .
\end{eqnarray}
To compute $\widetilde{K}_{m}$, one has to integrate by parts
and expand the integrand to the second order in $\theta$. The result is:
\begin{equation}
\label{B36a}
\widetilde{K}_m = -\frac{K_0}{2}+
\frac{3m^2}{4\pi\tau} 
\ . 
\end{equation}
We then find
\begin{eqnarray}
\label{B37}
\gamma_0^d&=&\eta_0
(N^2-1) (\widetilde{N} - \zeta),\nonumber\\[5pt]
\gamma_0^w&=&3\eta_0,\nonumber\\
\lambda^w&=&3N\eta_0 ,
\end{eqnarray}
Using Eqs.~(\ref{B31}), (\ref{C4}), (\ref{C6}) and~(\ref{C7}),  
we thus obtain, in the limit of small angle scattering and $|b_0| \tau \gg 1$,
\begin{equation}
\label{C9} \sigma_{\rm{sH}} = - \frac{e N}{4 \pi } \left(\frac{N^2 -
1}{N^2+1}\right) \left( \widetilde{N} - \zeta -2 \right) .
\end{equation}

As expected, this result gives $\sigma_{\rm{sH}} = 0$ for $|N| = 1$.
For $|N| \neq 1$, the result for small angle scattering can be larger
or smaller than (\ref{C8}), and can even have opposite sign, depending on
the values  of $\widetilde{N}$ and $\zeta$.

For a 2D hole system in GaAs, at low doping, with $N=\widetilde{N} = 3$
and $\zeta=0$, we obtain, for small angle scattering,
\begin{equation}
\label{C10} \sigma_{\rm{sH}} =  - \frac{3}{5}  \frac{e }{ \pi} ,
\end{equation}
which has 
different sign and is approximately twice as large as
the result (\ref{C8}) for isotropic scattering.

\begin{figure}[t]
\resizebox{.43\textwidth}{!}{\includegraphics{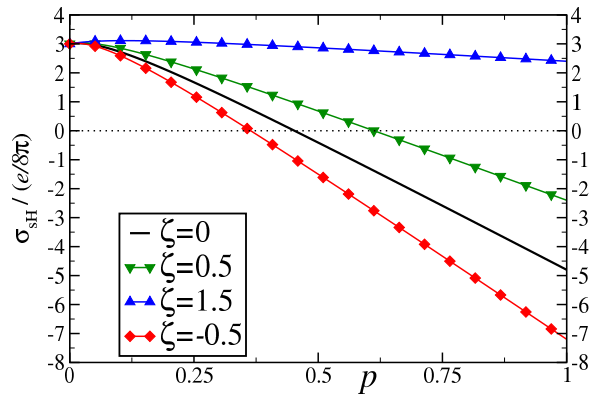}}
\vskip8mm
\caption{
Spin-Hall conductivity $\sigma_{{\rm sH}}$, in units
of $e/8\pi$, for various values of the band curvature parameter $\zeta$,
as a function of $p$, the fraction of transport scattering rate 
due to small-angle scatterers.
All curves are for parameters $N = \widetilde{N} = 3$.
(See Eqs. (3) and (4) for definitions.)
The clean limit, ($b_0 \tau \gg 1$), is assumed.
The case $\zeta = 0$ corresponds to a circularly-symmetric model
of a 2D hole gas in the limit of low carrier density.
}
\label{fig1}
\end{figure}

In a remotely-doped 2D electron or hole system, the charged impurities that 
compensate the carriers are located far from the 2D system, so they contribute only
long-wavelength Fourier components to the disorder potential, and
produce only small angle scattering.  As the set back distance is made  large,
their contribution to the transport scattering rate  decreases, 
so that eventually the latter may be dominated by a small
concentration of residual impurities close to the layer, whose scattering 
amplitudes may be nearly isotropic.  Thus, it is natural to consider a
model where both types of impurities are important, so the scattering 
kernel $K(\theta)$  contains both a narrow peak at $\theta \approx 0$, 
and an isotropic $\theta$-independent part. If the two contributions are weighted so
that a fraction $p$ of the transport scattering rate 
arises from small angle scatterers and a fraction
$1-p$ arises from large angle scatterers, then  doing the Fourier transform,
one finds that the coefficients $k_m$ 
entering (\ref{C7}) are given by $k_m = 1 - p +  p m^2$, for $m \neq 0$. 
[From (\ref{B21}) and (\ref{C4}), it may be seen that for 
arbitrary  $K(\theta)$, the transport scattering rate is 
determined by $K_0 - K_1$, so that one always has $k_1 = 1$.]
The constants $\gamma_0^d, \gamma_0^w,$ and $ \lambda^w,$ given 
by Eqs.~(\ref{B32})-(\ref{B34}) have no contribution from the isotropic part, and
are therefore each reduced from the values given in (\ref{B37}) by a factor of $p$. 
The quantity $\eta_0$ is independent of $p$. Substituting these values 
in the Eq.~(\ref{C6}) and~(\ref{C7}), one finds, in the clean limit, 
for $N \neq 1$:
\begin{eqnarray}
\label{sh-p}
\sigma_{{\rm sH}}  = \frac{e}{8\pi}\frac{N}{1 + pN^2} 
&\times& \Bigl\{
    (1 + 3p)  (1 - p )
\\
&-&    \left.p (N^2 - 1) \left[3p - 1 + 2p (\widetilde{N} - \zeta - 3) \right]
\right\} 
\ .
\nonumber 
\end{eqnarray}
Eq.~(\ref{sh-p}) reproduces the ``universal intrinsic value''~(\ref{C8}) 
for $p = 0$, and the small-angle scattering 
limit~(\ref{C9}) for $p=1$. 
Note, however, that it does not simply 
interpolate between the two values: Eq.~(\ref{sh-p}) is not necessarily  
a monotonic function of $p$.

In general, the parameters $\tilde{N}$ and $\zeta$ are related to each
other. However, they enter Eq.~(\ref{sh-p}) only via the combination
$\tilde{N} - \zeta$. 
In order  to illustrate the behavior of our model for different values of 
$\widetilde{N} - \zeta$, we plotted $\sigma_{sH} (p)$ at different values
of $\zeta$, keeping $\widetilde{N}$ fixed (see Fig.~\ref{fig1}). 
Note that spin-Hall conductivity can even change its sign 
for $\widetilde{N} - \zeta > 2$. Also, $\sigma_{sH}$ can exceed the
``universal intrinsic value'' $3e/8\pi$.

The model with a combination of  isotropic and small angle scattering
can also be considered as an approximation to a situation where there
is a single type of impurity with a scattering kernel that is neither isotropic
nor strongly peaked at small angles.
 
\section{Relation to experiments and previous theoretical results.}

In agreement with previous theoretical work, we have found  that
there is no spin Hall conductivity in the case $N= \pm 1 $,
applicable to a 2D electron system with pure Rashba or $\bk$-linear
Dresselhaus coupling, for arbitrary form of the impurity scattering,
while  the spin Hall conductivity has the full intrinsic value for
the case $N=3$, which models a 2D hole system,  when the impurity
scattering is isotropic.  For angle-dependent scattering, with
$N=3$, we find that the spin Hall conductivity is altered, so that
its precise value, and even its sign, can depend on such details as
the ratio of small angle to large angle scattering, and the energy
dependences of the spin-orbit coupling and of the hole velocity.
For the 2D hole systems that actually occur in GaAs,
an accurate computation should also take into account  warping of
the Fermi surface due to tetragonal anisotropy, which has been
omitted from our model.

Our results imply that spin-Hall conductivity for $N\neq1$
is not universal. Quantization
of $\sigma_{{\rm sH}}$ in units of $e/8\pi$ is broken by small-angle
scattering processes even in the clean limit ($b_0\tau \gg 1$), 
when spin-orbit field is strong. In contrast, charge Hall conductivity 
in the strong field limit is insensitive to the details of disorder scattering.
This difference probably reflects the non-topological origin 
of spin-Hall conductivity. This conclusion is indirectly 
supported by the result of numerical Laughlin gauge flux experiment performed
in Ref.~[\onlinecite{Haldane05}] for $N = 1$. 

In a recent article, Liu and Lei \cite{Liu} have performed numerical
calculations of the spin Hall conductivity for a two-dimensional hole
model, with scattering from impurities set back 50 nm from the layer.
Their model has a quadratic energy spectrum, and $\bk^3$ spin-orbit
coupling, corresponding to our model with $N= \widetilde{N} =3, \zeta=0$,
and has densities varying from $10^{10}$ to $10^{12} cm^{-2}$. 
(The densities in the published version of 
Ref.~\onlinecite{Liu} were misstated by a factor
of $100$).  Their  numerical results for $\sigma_{\rm{sH}}$  range from
slightly smaller than the intrinsic value, at their lowest densities, to
roughly twice the intrinsic value at high densities, in all cases quite
different from the prediction of our Eq.~(\ref{C10}).  For high
densities, it appears that their model is in the regime where the
spin-orbit splitting $b/v_F$ is larger than the momentum transfer in a
scattering event, so that our formulas would not apply.  (The numerical
results actually coincide with a prediction for this regime in a new work
by Khaetskii~\cite{Khaetskii05}) However, we do not have a complete
understanding of the numerical results at lower densities.

Although we do not know the precise value  to be
expected for the spin Hall conductivity in the 2D hole samples
studied by Wunderlich et al.,\cite{Wunderlich04} it seems plausible
that the magnitude should be similar to the universal intrinsic
value for $N=3$. In recent work,  Nomura et al.\cite{Wunderlich05} 
have analyzed the geometry of the experiment 
in Ref.~[\onlinecite{Wunderlich04}] and conclude that the
amount of polarized light obtained is consistent in magnitude with
what might be expected to arise from the predicted intrinsic spin Hall
conductivity.
Nevertheless, more work seems necessary before one
can be confident that one has a complete understanding
of the origin of the the observed polarization in these experiments.
As has been noted by various authors, spin polarization near a
boundary depends on the boundary conditions, and may not be simply
related to  spin Hall currents away from the boundary.  As an
example, for the 2D {\it electron} system with pure Rashba coupling,
where $\sigma_{\rm{sH}} = 0$, one may still find spin polarization
in the $z$-direction near the lateral  boundaries of a sample if
spin-flip processes are strong at the boundary,\cite{Rashba05} or
if electrons can cross the boundary into a contact or a region with
different spin-orbit coupling\cite{Mishchenko04,Adagideli05}. In any case,
it is expected that the polarization near a boundary will be insensitive  to
processes that occur further from the boundary than a few times
the length scale for spin relaxation, and this length scale is quite short
for holes in GaAs. It seems unlikely that one will be able to obtain a
{\em quantitative} measure of the bulk spin Hall conductivity in such
samples based on
observations of spin accumulation near a boundary.

In an earlier theoretical work by two of the present
authors,\cite{Mishchenko03} a set of Boltzmann-like kinetic equations
was derived  for a matrix distribution, denoted
$\hat{f}_{\bf p} (\mathbf{x} , t)$, which contained spin information but was
integrated over the energy $\epsilon$.  The model was restricted
to the case of $\bk$-linear spin-orbit coupling, and impurities were not included.
The focus of Ref.~[\onlinecite{Mishchenko03}] was on the time evolution of distributions
that are non-uniform in space,  and the linear response to a time-dependent
perturbation in the absence of impurity scattering. The formalism developed there
cannot be directly applied to spin-dependent transport in the presence of impurities,
which is the focus of the present work.

\section{ACKNOWLEDGMENTS.}

This work was supported in part by NSF grants PHY 01-17795 and
DMR 02-33773.
Research of AVS is supported by U.S. DOE under contract DEAC 02-98 CH 10886.
We are very thankful to A. Khaetskii, who derived independently spin-Hall
conductivity for small angle scattering\cite{Khaetskii05}, and
discussions with whom helped to correct errors in the previous versions of
our papers\cite{Shytov05v1,Khaetskii05v1}.
The authors are also grateful to S. Y. Liu for calling their attention to
the existence of Ref.~\onlinecite{Liu} and for discussions of the model
used in that paper.
Discussions with E. I. Rashba are also gratefully acknowledged.

\appendix

\section{Derivation of scattering kernel and the electric field term }

\subsection{Scattering kernel}

We give here a derivation of the two terms $\bM^d$and $\bM^w$
in the scattering kernel (\ref{B17}), using an approach based on
Fermi's Golden Rule.

The term $\bM^d$ arises from the spin-orbit corrections 
to the density of final states.
Consider an unpolarized electron of energy $\epsilon$, traveling
in a direction $\theta'$, incident on an impurity located at a point $\br_0$.
(The incident electron may be considered to be an incoherent 50-50 mixture
of spin up and spin down states along any convenient axis of quantization.)
Suppose the electron is scattered into a direction $\theta$.
At the angle $\theta$, there are two possible momenta
$k_\epsilon + \delta k$,
with $\delta k = b(\theta) \sigma / 2 v_\epsilon$,
where $\sigma = \pm 1$ denotes spin parallel or antiparallel to $\bb(\theta)$.
As in the discussion following (\ref{B6}), the corresponding
densities of states, per unit $\delta \theta$, are given by
$(2 \pi)^{-1} (k_\epsilon + \delta k) / (v_\epsilon + \delta v)$,
and the difference between $\sigma = \pm 1$ is given by
$\delta \tilde{\nu}
  = (2 \pi)^{-1} (\partial / \partial \epsilon) (k_\epsilon b /
  v_\epsilon)$.
If we ignore possible differences in matrix elements for the two spin
states, the difference in transition rates is given by
$W(q) \delta \tilde{\nu} /2$, where $W(q)$ is defined in (\ref{B12}).
This leads to a net polarization of the scattered spin,
giving a contribution to $\partial \bPhi (\theta)/ \partial t$
equal to $\bM^d (\theta, \theta') \delta n(\theta')$,
where $\bM^d$ is given by (\ref{B18}).  Since spin is conserved
during the scattering event, we will also have a contribution to
$\partial \bPhi (\theta')/ \partial t$
equal to $-\bM^d (\theta, \theta') \delta n(\theta')$.

The term $\bM^w$ arises from spin-dependent momentum shifts 
in the matrix elements.
The momentum transferred during the scattering process 
depends on the spin orientation of both
the initial and final state.  Letting $\sigma'=\pm1$ distinguish
between initial spin parallel or antiparallel to $\bb (\theta')$,
we see that the actual momentum transfer is $q+\delta q$, where $q$
is given by Eq. (\ref{B12}) and
\begin{equation}
\label{deltaq}
\delta q = (1/4) \left[ b(\theta) \sigma + b(\theta') \sigma' \right]
                 (\partial q / \partial \epsilon )
\end{equation}
is due to spin-orbit interaction. 
As we are assuming that $b/E_F \ll 1$, and we only need
the scattering kernel correct to linear order in $b$,  we may consider
the effects of $b(\theta)$ and $b(\theta')$ separately, and add the
results at the end.  Suppose that $b(\theta)=0$.  Then
$\delta q = [b(\theta') \sigma'/4] (\partial q / \partial \epsilon )$.
The change in the squared matrix element for the scattering process
is given by $\delta W = (\partial W/\partial q) \delta q$,
which leads to  different  scattering rates for $\sigma' = \pm 1$.
The out-scattering process thus creates a net polarization
of the electrons remaining at angle $\theta'$, and gives
a contribution to $\partial \bPhi (\theta' )/\partial t)$ equal to
\begin{equation}
\label{contrib1}
-\frac{\bb(\theta')}
{4k_\epsilon v_\epsilon}
\tan\left(\frac{\theta-\theta'}
{2}\right)\frac{\partial K(\theta-\theta')}
{\partial\theta}  \delta n(\theta')    .
\end{equation}
Since spin is conserved in the scattering process, we must have
a contribution to
$\partial \bPhi (\theta)/\partial t$ which is the negative of (\ref{contrib1}).

Next consider the case where $b(\theta') = 0$.  Now $\delta q$
depends  on $\sigma$, so the scattering rate depends on which
of the two final states is involved. The difference in matrix elements
leads to a contribution to the polarization at $\theta$ given by
\begin{equation}
\label{contrib2}
\frac{\partial \bPhi (\theta)} {\partial t} =
\frac{\bb(\theta)}
{4k_\epsilon v_\epsilon}
\tan\left(\frac{\theta-\theta'}
{2}\right)\frac{\partial K(\theta-\theta')}
{\partial\theta}  \delta n(\theta')    .
\end{equation}
Again, since spin is conserved in the scattering process, we must have
a contribution to
$\partial \bPhi (\theta')/\partial t$ which is the negative
of (\ref{contrib2}). The sum of these four contributions give
the contributions proportional to
$\bM^w(\theta, \theta')  \delta n(\theta')$
in Eq. (\ref{B16}).

Although we have used Fermi's Golden Rule and the Born approximation
in deriving these results,
we expect that Eqs. (\ref{B18}) and (\ref{B19}) should hold more generally,
within the model given by the Hamiltonian (\ref{B1}), as long as
the impurities are sufficiently dilute.

\subsection{Electric field term, Eq. (\ref{B15e}) }

The direct effect of a  uniform electric field term in the Hamiltonian,
in an infinitesimal time interval  $\delta t$,  is  to displace  the
momentum distribution by an amount $\delta  \bk  = \bE e \delta t$, while
keeping fixed the orientations of the spins. To see the effect of this, it is
convenient to integrate over energy and the magnitude of the momentum,
and define a spin density at an angle $\theta$ by
\begin{equation}
{\bf{m}}(\theta)=\int_0^\infty \frac{kdk}{8\pi^2} \int dE  \, 
[\bsig_{\alpha \beta} n_{\beta \alpha} (\bk.\epsilon)] ,
\end{equation}
with $\bk$ oriented in the direction $\theta$.
In thermal equilibrium, $\mathbf{m}$ is given by
\begin{equation}
\mathbf{m}^{\rm {eq}}(\theta) = \bb (\theta) k_F / ( 8 \pi^2 v_F ) ,
\end{equation}
under the assumption that  $b$ and the temperature $T$ are small
compared to $E_F$.

The electric field term conserves the total spin,
but gives rise to a flow of $\mathbf{m}$ around the unit circle at a velocity
$\dot{\theta}  =  - E e k_F^{-1} \sin \theta $, for  $\bE$ along the $x$-axis.
Then, to lowest order in the electric field, we have
\begin{equation}
\label{dmdt}
\left. \frac{\partial \mathbf{m} (\theta)}{\partial t} \right |_E
= - \frac{\partial}{\partial \theta} ( \dot{\theta} \mathbf {m}^{\rm{eq}} )
=    \frac{Ee}{  8 \pi^2 v_F}    \frac{\partial}{\partial \theta}  (\bb\sin \theta) .
\end{equation}

By definition, $[\mathbf{m}(\theta) - \mathbf{m}^{\rm{eq}}(\theta)] =
\int  d \epsilon \bPhi ( \theta, \epsilon) $.  Moreover, it is clear
that $\bPhi=0$ for energies far from the Fermi energy, and for any
fixed value of $\theta$, we must have $\partial \bPhi / \partial t |_E
\propto -\partial f_0 / \partial \epsilon $.  Equation (\ref{B15e}) then follows from
(\ref{dmdt})

\section{Alternate derivation using Green's functions}

The kinetic equation~(\ref{B13})-(\ref{B19}) contains non-diagonal
elements of the spin density matrix and is therefore a quantum kinetic
equation, rather than a conventional Boltzmann equation, which contains
only occupation numbers.
In this section we show that this equation can be derived from the
microscopic Hamiltonian~(\ref{B1}) by
Green's function methods.
In our derivation, we shall mainly follow the route outlined in
Ref.~[\onlinecite{RammerSmith}].
As usual, we introduce one-particle retarded and
advanced Green's function $\hat{G}_{R}$ and $\hat{G}_A$, and Keldysh function
$\hat{G}_K$ describing non-equilibrium state of the system.
(Note that all Green's functions are operators in spin space.)
Treating  the disorder potential as a perturbation, one may write Dyson
equation for these function in a matrix notation:
\begin{equation}
\label{dyson} \left(\hat G^{-1}_0-\underline{\hat \Sigma} \right)
\underline{\hat G}=1, ~~ 
\underline{\hat G} = \left(\begin{array}{ll} \hat G^R & \hat G^K \\
0 & \hat G^A
\end{array} \right)
\ .
\end{equation}
Here the lower bar denotes matrix in Keldysh space, and
$\hat{G}_0^{-1} = i \partial_t - \hat{H}$. In this work, we neglect
localization effects, and consider only diagrams with non-crossing
impurity lines. This gives the standard approximation for the
self-energy part
\begin{equation}
\underline{\hat \Sigma} ({\bf x}, t, {\bf x}', t')
= \widetilde{W}({\bf x} - {\bf x}')
  \underline{\hat G} ({\bf x},t, {\bf x}', t')\ ,
\end{equation}
where
$\widetilde{W}({\bf x} - {\bf x}') = \langle V({\bf x}) V({\bf x}') \rangle$
is the disorder correlation function.

For a time-independent disorder and in the absence of electron-electron
and electron phonon interactions, the functions $\hat{G}_R$ and $\hat{G}_A$ are
independent of Keldysh function $\hat{G}_K$. In fact, they are Hermitian
conjugate to each other:
\begin{equation}
\hat{G}_A(E, {\bf k}) = \hat{G}_R^{\dagger} (E, {\bf k})
\ ,
\end{equation}
and it is sufficient to find only one of them. Since the functions
$\hat{G}_R$ and $\hat{G}_A$ are independent of the non-equilibrium state, they
depend only on ${\bf x} - {\bf x}'$ in a translationally invariant
system. Going to  the Fourier representation,  one finds for $\hat{G}_R$:
\begin{equation}
\hat{G}_R (E, {\bf k}) = \frac{1}{E - \epsilon_{\bf k}
                                  + \frac{1}{2} {\bm \sigma}
                                \cdot{\bf b} ({\bf k})
                                  - \hat{\Sigma}_R (E, {\bf k})
                         }
\ ,
\end{equation}
 where $\epsilon_k$ is the hole dispersion without spin-orbit, and
\begin{equation}
\label{selfenergy}
\hat{\Sigma}_R(E, {\bf k}) = \int \frac{d^2{\bf k}'}{(2\pi)^2}
                                  W({\bf k} - {\bf k}')
                                  \hat{G}_R (E, {\bf k}')
\ ,
\end{equation}
where $W({\bf k})$ is the Fourier transform of the disorder
correlation function $\widetilde{W}({\bf x} - {\bf x}')$.

In general, the self-energy part~(\ref{selfenergy}) is an operator in
spin space. Its Hermitian part describing disorder-induced
renormalization of electron spectrum and spin-orbit coupling
will be ignored in the following.
The anti-Hermitian part of $\hat{\Sigma}_R (E, {\bf k})$ which we denote
$-i/2\hat{\tau}$ describes the decay of one-particle state because of elastic
scattering. It is convenient to project the function
$G_R({\bf k}', E)$ onto spin eigenstates:
\begin{eqnarray}
\label{g-projection}
\hat{G}_R(E, {\bf k}') &=& \sum_{\beta = \pm}
                   \frac{1}{E - E_\beta ({\bf k}')
                              + \frac{i}{2 \tau_{\beta} (E, {\bf k}')}
                   }
\nonumber
\\
                   &\times&
                   \left(1
                     + \frac{\beta {\bm \sigma}\cdot{\bf b}({\bf k}')}{b({\bf k}')}
                   \right)
\end{eqnarray}
where
\begin{equation}
\label{eigenvals}
E_\beta ({\bf k}) = \epsilon_{\bf k} - \frac{\beta}{2} b({\bf k})
\end{equation}
are corresponding eigenvalues, and $\tau_{\beta} (E, {\bf k})$
are lifetimes. Since the imaginary part of $\hat{G}_R(E, {\bf k}')$ is peaked
near $E = E_\beta({\bf k}')$, one can replace it by a delta-function to get
the self-energy in the leading order in impurity concentration. Thus,
we find
\begin{eqnarray}
\label{tau-so}
\frac{1}{\hat{\tau} (E, {\bf k})}
   &=&  2 \pi \int \frac{d^2{\bf k}'}{(2\pi)^2}  \sum_{\beta}
      \left(1 +  \frac{\beta {\bm \sigma} \cdot {\bf b}({\bf k}')}{b({\bf k}')}
      \right)
\\
   &\times&   W({\bf k} - {\bf k}')
      \delta (E - E_\beta({\bf k}'))
\ .
\nonumber
\end{eqnarray}
In this work, we restrict ourselves by the first order corrections in
spin-orbit coupling.  Also, we will be primarily interested in the
decay time on the ``mass shell'':
\begin{equation}
E  = \epsilon_{\bf k} - \frac{{\bm \sigma} \cdot {\bf b}(\bf k)}{2}
\ ,
\end{equation}
Expanding Eq.~(\ref{tau-so}) to the first order
in ${\bf b}({\bf k'})$, one finds, on the mass shell,
\begin{eqnarray}
\label{tau-so-final}
\frac{1}{\hat{\tau}_{\bf k}}
      &=& 2\pi\int \frac{d^2 {\bf k}'}{(2\pi)^2} W ({\bf k} - {\bf k}')
        \left[
               \delta (\epsilon_{\bf k} - \epsilon_{{\bf k}'})
         \right.
         \\
      &-&
         \left.
               {\bm \sigma} \cdot ({\bf b}({\bf k}) - {\bf b}({\bf k}'))
               \delta' (\epsilon_{\bf k} - \epsilon_{{\bf k}'})
        \right]
\ .
\nonumber
\end{eqnarray}
Thus, in general, scattering rates are different for spin majority and
spin minority bands.

Having found the retarded function $\hat{G}_R (E, {\bf k})$, one can now
rewrite the remaining equation for $\hat{G}_K$ to make it uniform
in $\hat{G}_K$ (see Ref.~[\onlinecite{RammerSmith}] for details):
\begin{equation}
\label{gk-1}
\hat{G}_R^{-1} \hat{G}_K - \hat{G}_K \hat{G}_A^{-1}
= \hat{G}_R \hat{\Sigma}_K - \hat{\Sigma}_K \hat{G}_A
\ ,
\end{equation}
where
\begin{equation}
\hat{\Sigma}_K({\bf x}, {\bf x}', t, t') = \widetilde{W} ({\bf x} - {\bf x}')
\hat{G}_K ({\bf x}, {\bf x}', t, t')
\ ,
\end{equation}
and operators $\hat{G}_R^{-1}$ and $\hat{G}_A^{-1}$ act on the left and
the right argument of $\hat{G}_K$ respectively. We now apply Wigner
transform:
\begin{eqnarray}
\hat G^K(t,{\bf x};t',{\bf x}') &=& \int
\frac{dE}{2\pi}\,\frac{d^2{\bf k}}{(2\pi)^2}\,
                             e^{i{\bf k}\cdot({\bf x} - {\bf x}')-iE (t-t')}
                             \nonumber
			     \\
                            &\times&
			    \hat{g}_{{\bf k}E}
                            \left(
                                  \frac{t + t'}{2},
                                  \frac{{\bf x}+ {\bf x}'}{2}
                            \right)
\ , 
\end{eqnarray}
and the left hand side of Eq.~(\ref{gk-1}) becomes
\begin{eqnarray}
\label{gk-2}
& &  (\hat{G}_R^{-1} \hat{G}_K - \hat{G}_K \hat{G}_A^{-1})
\\
&=&
  \frac{\partial \hat{g}_{{\bf k}E}}{\partial t}
+ \frac{1}{2}\left\{\frac{{\bf k}}{m} - \frac{1}{2}
                 \frac{\partial ({\bf b}\cdot{\bm \sigma})}{\partial {\bf k}} ,
        {\bm \nabla}\hat{g}_{{\bf k}E}\right\}
\nonumber
\\
&-&  \frac{i}{2} \left[ {\bm \sigma} \cdot {\bf b}({\bf k}),
                     \hat{g}_{{\bf k}E} \right]
+ \frac{1}{2} \left\{ \frac{1}{\hat{\tau}_{\bf k} },
                      \hat{g}_{{\bf k}E}\right\}
\ ,
\nonumber
\end{eqnarray}
where the scattering rate operator $1/\hat{\tau}_{\bf k}$ is given by
Eq.~(\ref{tau-so-final}).  The right hand side of Eq.~(\ref{gk-1})
takes the form
\begin{eqnarray}
\label{gk-3}
& & ( \hat{G}_R \hat{\Sigma}_K - \hat{\Sigma}_K \hat{G}_A )
\\
&=& \int \frac{d^2 {\bf k}'}{(2\pi)^2} W({\bf k} - {\bf k}')
   \left[
        \hat{G}_R (E, {\bf k}) \hat{g}_{{\bf k}' E} ({\bf x}, t)
    \right.
\nonumber
\\
      &-&
     \left.
     \hat{g}_{{\bf k}' E} ({\bf x}, t) \hat{G}_A (E, {\bf k})
   \right]
\nonumber
\end{eqnarray}
(here we assumed that ${\bf x}$- and $t$- dependence of
$\hat{g}_{{\bf k}E}$ is smooth on the scale of Fermi wavelength).

Calculating transport quantities, we are normally interested in
Keldysh Green's function at coinciding temporal points, $t = t'$,
which in Wigner representation corresponds to energy-integrated
functions.
However, Eqs.~(\ref{gk-1})-(\ref{gk-3}) contains Green's function
for arbitrary energies. To express all quantities in terms of
functions at $t=t'$, we use the following ansatz for
$\hat{g}_{{\bf k}E}$:
\begin{equation}
\label{h-def}
\hat{g}_{{\bf k}E} = \hat{G}_R (E, {\bf k}) \hat{h}_{{\bf k}} -
                     \hat{h}_{{\bf k}}  \hat{G}_A (E, {\bf k})
\ ,
\end{equation}
where the function $\hat{h}_{\bf k}$ depends only on
momentum. Integrating Eq.~(\ref{h-def}) over $E$, one finds:
\begin{equation}
\int \frac{dE}{2\pi} \hat{g}_{{\bf k}E}  = i \hat{h}_{\bf k}
\ .
\end{equation}
Therefore, the function $\hat{h}_{\bf k}({\bf x}, t)$ can be used to compute
transport quantities. In fact, the function
$\hat{f}_{{\bf k}} = 1 - 2 \hat{h}_{\bf k}$
can be interpreted as a distribution function, with diagonal elements
of $\hat{f}_{{\bf k}}$ being the  occupation numbers.

Now we can derive the equation for the function
$\hat{h}_{\bf k} ({\bf x}, t)$
by integrating left- and right-hand sides of quantum kinetic
equation~(\ref{gk-2}) and~(\ref{gk-3}). Since the left-hand side
depend on $E$ only via $\hat{G}_{K}(E)$, the integration is trivial.
The terms on the right hand side (Eq.~(\ref{gk-3})) contain $E$ explicitly,
and should be carefully integrated. Substituting the
definition~(\ref{h-def}) into the Eq.~(\ref{gk-3}), one gets four
different terms. Two of them contain the product of two retarded (or
two advanced) Green's functions, and vanish after integration
over $E$. The two remaining terms are anti-Hermitian conjugate to each
other. One of them has the form
\begin{equation}
R_1  = \int \frac{dE}{2\pi}
                     \hat{G}_R (E, {\bf k}) \hat{h}_{{\bf k}'}
                     \hat{G}_A (E, {\bf k}')
\ .
\end{equation}
Projecting Green's function onto eigenstates with the help of
Eq.~(\ref{g-projection}) and integrating over $E$, one finds:
\begin{eqnarray}
R_1 &=& i \sum_{\beta, \beta' = \pm}
        \left(
         1 + \beta \frac{{\bm \sigma}\cdot{\bf b}({\bf k})}{b({\bf k})}
    \right)
\\
\nonumber
&\times&  \frac{\hat{h}_{\bf k}'}{E_{\beta}({\bf k}) - E_{\beta'}({\bf k}')
      + \frac{i}{2}\left(\frac{1}{\tau_\beta({\bf k})}
                       + \frac{1}{\tau_{\beta'}({\bf k}')}
                            \right)}
\\
&\times&
       \left(1 + \beta' \frac{{\bm \sigma}\cdot {\bf b}({\bf k}')}{b({\bf k}')}
\right)
\nonumber
\ ,
\end{eqnarray}
where the spin-majority and spin-minority energy bands are given by
the Eq.~(\ref{eigenvals}).  Assuming the scattering rate $1/\tau_\sigma$
in the denominator to be small, we now rewrite this term using
Sokhotsky formula:
\begin{equation}
\frac{1}{x + i 0} = \mathop{\rm P}\nolimits\frac{1}{x}  - i \pi \delta (x)
\ ,
\end{equation}
where ${\rm P}$ denotes principal value. Subtracting Hermitian
conjugate and expanding to the first order in spin-orbit field
${\bf b}({\bf k})$, one finds, for the right hand side of quantum kinetic
equation:
\begin{eqnarray}
\label{terms}
R_1 - R_1^{\dagger} &=&
      2\pi \hat{h}_{{\bf k}'} \delta (\epsilon_{{\bf k}}
                                      - \epsilon_{{\bf k}'})
\\
 &-&   \pi
       \left\{\hat{h}_{{\bf k}'},
                {\bm \sigma} \cdot
            ({\bf b} ({\bf k}) - {\bf b}({\bf k}'))\right\}
       \delta' (\epsilon_{{\bf k}} - \epsilon_{{\bf k}'})
\nonumber
\\
 &-&    i \mathop{{\rm P}}\nolimits\frac{\left[\hat{h}_{{\bf k}'},
                {\bm \sigma} \cdot
               ({\bf b} ({\bf k}) + {\bf b}({\bf k}'))\right]
          }{(\epsilon_{{\bf k}} - \epsilon_{{\bf k}'})^2}
\nonumber
\ .
\end{eqnarray}
The first term in the Eq.~(\ref{terms}) is a conventional scattering
term, while the second is the spin-orbit correction to the scattering
rate. The third term is due to renormalization of one-particle
spectrum by the disorder, and should be ignored in the following,
since we already neglected similar terms in the real part of
self-energy.
Now we can use the Eq.~(\ref{terms}) and~(\ref{gk-2}) to
write the equation for $\hat{h}_{{\bf k}}$:
\begin{eqnarray}
& & \frac{\partial \hat{h}_{{\bf k}}}{\partial t}
+ \frac{1}{2}\left\{\frac{{\bf k}}{m} - \frac{1}{2}
                              \frac{\partial ({\bf b}\cdot{\bm \sigma})}{\partial {\bf k}} ,
        {\bm \nabla}\hat{h}_{{\bf k}}\right\}
\\
&-&  \frac{i}{2} \left[ {\bm \sigma}\cdot {\bf b}({\bf k}),
                     \hat{h}_{{\bf k}} \right]
\nonumber
\\
&=& 2\pi \int \frac{d^2{\bf k}'}{(2\pi)^2} W({\bf k} - {\bf k}')
\left\{
(\hat{h}_{{\bf k}'} - \hat{h}_{{\bf k}}) \delta (\epsilon_{\bf k}'
                                          - \epsilon_{{\bf k}'})
\right.
\nonumber
\\
&-&
\left.
\frac{1}{2}
\left\{\hat{h}_{{\bf k}'} - \hat{h}_{{\bf k}} ,
                 {\bm \sigma} \cdot
                ({\bf b} ({\bf k}) - {\bf b}({\bf k}'))\right\}
       \delta' (\epsilon_{{\bf k}} - \epsilon_{{\bf k}'})
\right\}
\nonumber
\end{eqnarray}
(we grouped together terms proportional to $W$).

To this point, we did not take into account the effects of external electric
field. To incorporate the electric field into kinetic equation,
consider a very smooth external scalar potential $U({\bf r})$.
Strictly speaking, this potential breaks translational invariance,
and the calculations we have done so far become invalid.
However,  for small field the potential $U({\bf r})$ is smooth
on the spin-orbit precession length $\hbar v_\epsilon / b$ 
and does not lead to transitions between spin subbands.
Therefore, $U({\bf r})$ can be considered as a local shift of electron energy.
The term  $\hat{G}_R^{-1} \hat{G}_K - \hat{G}_K \hat{G}_A^{-1}$
now contains $U({\bf r}) - U({\bf r}')$, which becomes, after
Wigner transform, $\nabla_{{\bf r}} U  \nabla_{\bf k} \hat{g}_{{\bf k}E}$.
Thus, the electric field term in the kinetic equation takes the
standard  form.  Rewriting kinetic equations in terms of function
$\hat{f}_{{\bf k}} = 1 - 2 \hat{h}_{\bf k}$, one finds:
\begin{eqnarray}
\label{ke-momentum-final}
& & \frac{\partial \hat{f}_{{\bf k}}}{\partial t}
+ \frac{1}{2}\left\{\frac{{\bf k}}{m} - \frac{1}{2}
                              \frac{\partial ({\bf b} \cdot {\bm \sigma})}{\partial {\bf k}} ,
        {\bm \nabla}\hat{f}_{{\bf k}}\right\}
\\
&-&  \frac{i}{2} \left[ {\bm \sigma} \cdot {\bf b}({\bf k}),
                     \hat{f}_{{\bf k}} \right]
 + e{\bf E} \frac{\partial \hat{f}_{{\bf k}}}{\partial {\bf k}}
\nonumber
\\
&=& 2\pi \int \frac{d^2{\bf k}'}{(2\pi)^2} W({\bf k} - {\bf k}')
\Bigl(
(\hat{f}_{{\bf k}'} - \hat{f}_{{\bf k}}) \delta (\epsilon_{\bf k}'
                                               - \epsilon_{{\bf k}'})
\nonumber
\\
&-&
\frac{1}{2}
\left\{\hat{f}_{{\bf k}'} - \hat{f}_{{\bf k}} ,
                 {\bm \sigma} \cdot
               ({\bf b} ({\bf k}) - {\bf b}({\bf k}'))\right\}
       \delta' (\epsilon_{{\bf k}} - \epsilon_{{\bf k}'})
\Bigr)
\nonumber
\ .
\end{eqnarray}

The kinetic equation~(\ref{ke-momentum-final}) is sufficient to
compute spin transport in a 2D hole gas. Note, however, that in
general, spin-orbit terms on the right hand side mix states with
different values of $\epsilon_{\bf k}$ because of the presence
of the derivative of delta-function. 
Thus, in general, 
one has to consider all electron states with different energies
simultaneously. This mixing occurs
because the quantity $\epsilon_{\bf k}$ is not conserved in the
scattering process. Instead, the true electron energy
\begin{equation}
\epsilon = \epsilon_{\bf k} - \frac{{\bm \sigma} \cdot {\bf b}({\bf k})}{2}
\end{equation}
is conserved. We are now going to use this integral of motion
to separate contribution of electrons with different energies,
and make energy conservation explicit.
We shall make the change of variables, expressing the electron
momentum in terms of energy $\epsilon$ and momentum direction
$\theta$. Note, however, that the electron energy $\epsilon$
contains spin-dependent part, and therefore replacing momentum
length by $\epsilon$ is not a simple algebraic transformation:
one has to ensure the proper ordering of spin operators in the resulting
expression. To figure out this operator order, consider the effect of
spin-dependent gauge transformation
\begin{equation}
\label{u-transform}
\Psi'({\bf r}) = \hat{U}_\sigma \Psi({\bf r})
= \exp (i ({\bf q}\cdot{\bf r}) ({\bm \sigma}\cdot{\bf b})) \Psi({\bf r})
\end{equation}
on the one-particle density matrix
$\hat{\rho}_{{\bf r}, {\bf r}'}
 = \Psi^{\dagger}({\bf r}) \Psi({\bf r}')$.
In the first order in ${\bf b}$, one finds:
\begin{eqnarray}
\hat{\rho}^{'}_{{\bf r},{\bf r}'}
&=& \hat{U}_\sigma^{\dagger} \hat{\rho}_{{\bf r},{\bf r}'}\hat{U}_{\sigma}
    \approx \hat{\rho}_{{\bf r},{\bf r}}
    - \frac{i {\bf q}\cdot ({\bf r} + {\bf r}')}{2}
      \left[\hat{\rho}_{{\bf r},{\bf r}'}, {\bm \sigma}\cdot{\bf b}\right]
\nonumber
\\
&-& \frac{i {\bf q} \cdot ({\bf r} - {\bf r}')}{2}
      \left\{\hat{\rho}_{{\bf r},{\bf r}'}, {\bm \sigma}\cdot{\bf b}\right\}
\end{eqnarray}
By doing Wigner transform with respect to ${\bf r}$ and ${\bf r}'$,
one arrives at:
\begin{eqnarray}
\label{rho-transform}
\hat{\rho}'({\bf R}, {\bf p}) &=& \hat{\rho}({\bf R}, {\bf p})
- i  \left[ \hat{\rho} ({\bf R}, {\bf p}) , ({\bf q}\cdot{\bf R})
                                            ({\bm \sigma}\cdot{\bf b})\right]
\nonumber
\\
&+& \left\{ ({\bf q}\cdot \nabla_{\bf p}) \hat{\rho}({\bf R}, {\bf p}) ,
            ({\bm \sigma}\cdot{\bf b})\right\}
\ .
\end{eqnarray}
The second term in the Eq.~(\ref{rho-transform}) describes spin
rotation due to transformation $\hat{U}_{\sigma}$, and the third term
describes spin-dependent momentum shift. Note that spin operator in
the last term acts symmetrically on $\hat{\rho}$. This symmetry is related
to  the antisymmetric action of momentum operator on $\Psi ({\bf r})$ and
$\Psi^{\dagger} ({\bf r}')$ in $\rho_{{\bf r}, {\bf r}'}$. One can
undo the spin rotation described by the second term in the
Eq.~(\ref{rho-transform}), and use the third term as a prescription
of how the spin-dependent momentum shift should be performed:
\begin{equation}
\label{prescription}
\hat{\rho} ({\bf R}, {\bf p} + {\bf q} ({\bm \sigma}{\bf b}))
= \hat{\rho} ({\bf R}, {\bf p})
+ \frac{1}{2} \left\{ {\bf q}{\bm \nabla}_{\bf p} \hat{\rho} ({\bf R}, {\bf p}),
          ({\bm \sigma}{\bf b})  \right\}
\ .
\end{equation}

Now we can replace momentum in the Eq.~(\ref{ke-momentum-final}) by
\begin{equation}
\label{momentum-transform}
{\bf k} = k_\epsilon \hat{k}
        + \frac{{\bm \sigma}\cdot{\bf b}}{2v_\epsilon} \, \hat{k}
\ ,
\end{equation}
where $\hat{k} = (\cos\theta, \sin\theta)$ is the unit vector in
the momentum direction, and $k_\epsilon$ and
$v_\epsilon = \partial \epsilon_k / \partial k$ are defined
in Sec.~\ref{boltzmann-formulation}. To rewrite the gradient term
in this representation, one can simply replace the momentum
according to the Eq.~(\ref{momentum-transform}):
\begin{equation}
\label{term-g}
\left.\frac{\partial \hat{f}}{\partial t}\right|_{v}
=  - \frac{1}{2} \left\{ v_\epsilon \hat{k}
                      + \frac{\partial v_\epsilon}{\partial \epsilon}
                         \frac{{\bm \sigma}\cdot{\bf b}}{2} \hat{k}
                      -
                \frac{v_\epsilon}{2}
            \frac{\partial ({\bf b} ({\bf k})\cdot{\bm\sigma})}{
                                          \partial \epsilon},
                         \nabla_{{\bf r}} \hat{f}
               \right\}
\ .
\end{equation}
The spin-orbit rotation term already contains spin-orbit field
and should not be modified at all:
\begin{equation}
\label{term-b}
\left.\frac{\partial \hat{f}}{\partial t}\right|_{b} = -
\frac{i}{2} \left[ {\bm \sigma} \cdot {\bf b}({\bf k}), \hat{f} \right]
\ .
\end{equation}
To find the electric field contribution, note that for small field
it should be computed only for an equilibrium distribution.
In the equilibrium, the occupation numbers $\hat{f}_{\bf k}$ depend only
on the total energy. Thus,
\begin{eqnarray}
\label{term-e}
\left.\frac{\partial \hat{f}}{\partial t}\right|_{E}
&=& eE \frac{\partial f_0}{\partial \epsilon}
\frac{\partial \epsilon}{\partial k_x}
= eE \frac{\partial f_0}{\partial \epsilon}
\left[ v_\epsilon (k_\epsilon) \cos\theta
\right.
\\
&+&
\frac{{\bm \sigma}\cdot{\bf b}(k_\epsilon)}{2}
  \frac{\partial v_\epsilon}{\partial \epsilon} \cos\theta
- \frac{v_\epsilon}{2}\frac{\partial ({\bf b}\cdot{\bm \sigma})}{\partial \epsilon}
             \cos\theta
\nonumber
\\
&+&
\left.
      \frac{1}{2}\frac{\partial ({\bf b}\cdot{\bm \sigma})}{\partial \theta}
       \frac{\sin\theta}{k_\epsilon}\right]
\ .
\nonumber
\end{eqnarray}
(the second term here takes into account the momentum shift given by
Eq.~(\ref{momentum-transform})).

In the scattering term, the delta-function and its derivative
combine into $\delta(\epsilon - \epsilon')$, due to energy conservation.
In zeroth order, the scattering contribution is
\begin{equation}
\label{scattering-0}
\left.\frac{\partial \hat{f}}{\partial t}\right|_{0}
= \int \frac{k_\epsilon d\theta'}{2\pi v_\epsilon}
  W({\bf k} - {\bf k}') \left( \hat{f}(\theta') -
  \hat{f}(\theta)\right)
\ ,
\end{equation}
where both ${\bf k}$ and ${\bf k}'$ are taken on the surface
$\epsilon_{k} = \epsilon$.
Spin-orbit corrections to the Eq.~(\ref{scattering-0})
arise  from either volume element
$d^2{\bf k}'$ or the matrix element
$W({\bf k} - {\bf k}')$. The matrix element contribution is
due to spin-orbit induced changes of ${\bf k}$ and ${\bf k}'$:
\begin{eqnarray}
\label{term-w}
& &\left.\frac{\partial \hat{f}}{\partial t}\right|_{w} = -
\int \frac{k_\epsilon d\theta'}{8\pi v_\epsilon^2}
\\
&\times&
\left\{
   {\bm \sigma} \cdot\left(\frac{\partial W}{\partial k} {\bf b}({\bf k})
 + \frac{\partial W}{\partial k'} {\bf b}({\bf k'})
   \right),
   \hat{f}(\epsilon, \theta') - \hat{f}(\epsilon, \theta)
\right\}
\ .
\nonumber
\end{eqnarray}
Transforming this expression as explained in Appendix A, one
recovers the scattering kernel ${\bf M}_w$.

The volume element is computed as follows:
\begin{equation}
\label{volume-element}
d^2{\bf k} = k dk d\theta =
 \left(k_\epsilon + \frac{{\bm \sigma}\cdot {\bf b}}{2v_\epsilon}\right)
 \frac{dk}{d\epsilon} d\theta
\ .
\end{equation}
The derivative $dk/d\epsilon$ should be computed to the first order in
spin-orbit field, as explained in comments to Eq.~(\ref{B6}), and
the volume element becomes
\begin{equation}
d^2{\bf k} = d\epsilon d\theta
\left(\frac{k_\epsilon}{v_\epsilon}  + \frac{1}{2}
\frac{\partial}{\partial \epsilon}
           \frac{k_\epsilon {\bf b}_\epsilon \cdot {\bm \sigma}}{v_\epsilon}
 \right)
\end{equation}
(as usual, its matrix part should be applied symmetrically).
This gives additional contribution to the collision integral:
\begin{eqnarray}
\label{term-d}
\left.\frac{\partial \hat{f}}{\partial t}\right|_{d}
&=& - \int \frac{d\theta'}{4\pi}
W ({\bf k} - {\bf k}')\,
\\
&\times&
\left\{
     \frac{\partial}{\partial \epsilon}
           \frac{k_\epsilon {\bm \sigma}\cdot{\bf b}({\bf k}_\epsilon')}{
             2 v_\epsilon},
     \hat{f}(\epsilon, \theta') - \hat{f}(\epsilon, \theta)
\right\}
\ .
\nonumber
\end{eqnarray}
The evolution of the density matrix $\rho$ in $\epsilon-\theta$
representation is then given by
\begin{equation}
\label{eq-abstract}
\frac{\partial h}{\partial t} =
\left.\frac{\partial \hat{f}}{\partial t}\right|_{g}
+  \left.\frac{\partial \hat{f}}{\partial t}\right|_{b}
+  \left.\frac{\partial \hat{f}}{\partial t}\right|_{E}
+  \left.\frac{\partial \hat{f}}{\partial t}\right|_{0}
+  \left.\frac{\partial \hat{f}}{\partial t}\right|_{d}
+  \left.\frac{\partial \hat{f}}{\partial t}\right|_{w}
\ ,
\end{equation}
where all terms are given by Eqs.~(\ref{term-g}),
(\ref{term-b}), (\ref{term-e}), (\ref{term-w}) and~(\ref{term-d}).

To simplify calculation of physical quantities, one can remove the
spin dependence from the volume element. To do that, we  include the
volume element in the distribution function via the following transformation:
\begin{equation}
\label{n-subst}
\hat{n}(\epsilon, \theta)
= \frac{k_\epsilon}{v_\epsilon} \hat{f}(\epsilon, \theta)
+ \frac{1}{4}\left\{\hat{f}(\epsilon, \theta),
      \frac{\partial}{\partial \epsilon}
        \frac{k_\epsilon {\bm \sigma} \cdot {\bf b} ({\bf k}_\epsilon)}{v_\epsilon}
   \right\}
\ .
\end{equation}
One can see that in the first order in spin-orbit
\begin{equation}
\frac{1}{2} \left\{ \hat{f}(\epsilon, \theta),  d^2{\bf k} \right\}
= \hat{n}(\epsilon, \theta) d\epsilon d\theta
\ .
\end{equation}
The function $\hat{n}(\epsilon, \theta)$ can be now identified
with the distribution used throughout the paper. Substituting
Eqs.~(\ref{n-subst})  and~(\ref{B7})
into Eqs.~(\ref{term-e}), (\ref{term-w})  and~(\ref{term-d}),
one can reproduce Eqs.~(\ref{B15e}),
(\ref{B19}), and~(\ref{B18}) respectively.
Indeed, the spin-orbit correction to the electric field term
is given by the sum of contributions from Eq.~(\ref{term-e})
and Eq.~(\ref{n-subst}):
\begin{eqnarray}
\left.\frac{\partial \hat{n}}{\partial t}\right|_{E,so}
&=& -eE \frac{\partial f_0}{\partial \epsilon} \frac{{\bm \sigma}}{2}
\cdot \left\{
     v_\epsilon \cos\theta
     \frac{\partial}{\partial \epsilon}
     \frac{k_\epsilon {\bf b}_\epsilon}{v_\epsilon}
\right.
\\
&+& \frac{k_\epsilon {\bf b}_\epsilon}{v_\epsilon} \cos\theta
     \frac{\partial v_\epsilon}{\partial \epsilon}
 - k_\epsilon \cos\theta \frac{\partial {\bf b}_\epsilon}{\partial \epsilon}
\nonumber
\\
&+&
 \left.
 \frac{\sin\theta}{v_\epsilon} \frac{\partial {\bf b}}{\partial \theta}
\right\}
\nonumber
\ .
\end{eqnarray}
Collecting the terms proportional to $\cos\theta$, one finds:
\begin{equation}
\left.\frac{\partial \hat{n}}{\partial t}\right|_{E,so}
=  eE \frac{\partial f_0}{\partial \epsilon} \frac{{\bm \sigma}}{2}
\left\{
\cos\theta \frac{{\bf b}_\epsilon}{v_\epsilon}
- \frac{\sin\theta}{v_\epsilon} \frac{\partial {\bf b}}{\partial \theta}
\right\}
\ ,
\end{equation}
thus reproducing Eq.~(\ref{B15e}). The matrix element correction
to the collision integral~(\ref{term-w}) reproduces Eq.~(\ref{B19}).
The density of states correction~(\ref{term-d}), however, is modified
by a similar contribution that arises from transformation of the
collision integral~(\ref{scattering-0}). In the latter, one
has to express $\hat{f}(\epsilon, \theta')$ in terms of
$\hat{n}(\epsilon, \theta)$ and {\it then} transform the
resulting contribution according to Eq.~(\ref{n-subst}) to
get $\partial \hat{n}/\partial t$. As a result, four spin-orbit
terms arise.
Then, the term in Eq.~(\ref{term-d})
containing $\hat{n}(\theta')$  is cancelled by the corresponding
contribution to the Eq.~(\ref{scattering-0}). Similar cancellation
occur for $\hat{n}(\theta)$ term. As a result, one
recovers Eq.~(\ref{B18}).
This concludes
the derivation of kinetic equation~(\ref{B13}) by Green's function
method.


\begin{thebibliography}{99}

\bibitem{DP71} M.I. Dyakonov and V.I. Perel,  Phys.\ Lett.\ A {\bf 35}, 459 (1971).


\bibitem{Kato04}  Y.K. Kato, R.C. Myers, A.C. Gossard, and D.D. Aw\-scha\-lom,
                  Science {\bf 306}, 1910 (2004).

\bibitem{Wunderlich04}  J. Wunderlich, B. Kaestner, J. Sinova, and T. Jungwirth,
Phys.\ Rev.\ Lett.\ {\bf 94}, 047204  (2005).


\bibitem{Sih05} V. Sih, R. C. Meyers, Y. R. Kato, W. H. Lau, A.C. Gossard, and D. D. Awschalom,
cond-mat/0506704.

\bibitem{H99} J.E. Hirsch, Phys. Rev.
Lett.\ {\bf 83}, 1834 (1999).

\bibitem{Zhang00} S. Zhang, Phys. Rev. Lett.  {\bf 85}, 393 (2000).

\bibitem{Murakami03} S. Murakami, N. Nagaosa, and S.C. Zhang,
 Science {\bf 301}, 1348 (2003).

\bibitem{Sinova04} J. Sinova, D. Culcer, Q. Niu, N.A. Sinitsyn, T. Jungwirth,
and A.H. MacDonald, Phys.\ Rev.\ Lett.\ {\bf 92}, 126603 (2004).

\bibitem{Engel05} H.A. Engel, B.I. Halperin, and E.I. Rashba, E.I., cond-mat/0505535.

\bibitem{Tse05} W-K. Tse, S. Das Sarma, cond-mat/0507149.

\bibitem{Bernevig04} B.A. Bernevig, and S. Zhang, cond-mat/0412550.

\bibitem{Rashba03} E.I.~Rashba, Phys. Rev. B {\bf 68}, 241315(R)
(2003).

\bibitem{Schliemann04} J. Schliemann, and D. Loss, Phys.\ Rev.\ B {\bf 69},
165315 (2004).

\bibitem{Burkov04} A.A. Burkov, and L. Balents,
Phys. Rev. B {\bf 69}, 245312 (2004).

\bibitem{Inoue04} J.I. Inoue, G.E.W. Bauer, and L.W. Molenkamp, Phys.\ Rev.\ B, {\bf 70}, 041303(R) (2004).

\bibitem{Mishchenko04} E.G. Mishchenko, A.V. Shytov, and B.I. Halperin,
Phys.\ Rev.\ Lett.\ {\bf 93}, 226602 (2004).

\bibitem{Khaetskii04} A. Khaetskii, cond-mat/0408136.

\bibitem{Raimondi05} R. Raimondi, and P. Schwab, Phys. Rev. B {\bf 71}, 033311 (2005).

\bibitem{Rashba04} E.I. Rashba, cond-mat/0404723.

\bibitem{Murakami05} N. Sugimoto, S. Onoda, S. Murakami, and N. Nagaosa,
cond-mat/0503475.

\bibitem{Nomura04} K. Nomura, J. Sinova, T. Jungwirth, Q.Niu, 
                      and A. H. MacDonald 
                   Phys. Rev. B {\bf 71}, 041304(R) (2005).

\bibitem{Nomura05}  K. Nomura, J. Sinova, N.A. Sinitsyn, and A.M. MacDonald,
cond-mat/0506189.

\bibitem{Dimitrova04} O.V. Dimitrova, cond-mat/0405339, v. 2.

\bibitem{Vasko79} F.T.\ Vas'ko, JETP Lett.\ {\bf 30}, 540 (1979);
Yu.A.~Bychkov and E.I.~Rashba, J.~Phys.\ C {\bf 17}, 6039 (1984).


\bibitem{Chalaev04} O. Chalaev and D. Loss, 
                    Phys. Rev. B {\bf 71}, 245318 (2005).

\bibitem{Malshukov04} A.G. Mal'shukov, K.A. Chao, 
                      Phys. Rev. B {\bf 71}, 121308(R), 2005.

\bibitem{Murakami04} S. Murakami, Phys. Rev. B {\bf 69}, 241202(R) (2004).

\bibitem{AGD} A.A. Abrikosov, L.P. Gorkov, and I.E. Dzyaloshinskii,
{\it Methods of quantum field theory in statistical physics},
(Prentice-Hall, Englewood Cliffs, NJ, USA) 1963.

\bibitem{Nikolic1} B.K. Nikoli\'c, L.P. Z\^arbo, and  S. Souma, 
                   Phys. Rev. B {\bf 72}, 075361 (2005).

\bibitem{Hank04} E.M. Hankiewicz, L.W. Molenkamp, T. Jungwirth, and J. Sinova, 
		 cond-mat/0409334.

\bibitem{Nikolic2} B.K. Nikolic, S. Souma, L.P. Zarbo, and J. Sinova, 
                   Phys. Rev. Lett. {\bf 95}, 046601 (2005).

\bibitem{Wunderlich05}  K. Nomura, J. Wunderlich, J. Sinova, B. Kaestner,
A.H. MacDonald, and T. Jungwirth, cond-mat/0508532.

\bibitem{Rashba05} E.I. Rashba, cond-mat/0507007.

\bibitem{Adagideli05} I. Adagideli and G. E.W. Bauer,
                      cond-mat/0506531.

\bibitem{Haldane05}  D.N. Sheng, L. Sheng, Z.Y. Weng, F.D.M. Haldane, 
                     cond-mat/0504218.
		     
\bibitem{Liu} S.Y.~Liu and X.L.~Lei, Phys. Rev. B 72, 155314 (2005); 
              and erratum (in press).
	      
\bibitem{Mishchenko03} E.G. Mishchenko and B.I. Halperin,
                       Phys. Rev. B{\bf 68}, 045317 (2003).

\bibitem{Khaetskii05} A.~Khaetskii, cond-mat/0510815 v2 

\bibitem{Khaetskii05v1} A.~Khaetskii, cond-mat/0510815 v1

\bibitem{Shytov05v1} A.V.~Shytov, E.G.~Mishchenko, H.A.~Engel, 
                     and B.I.~Halperin, cond-mat/0509702 v1

		       
\bibitem{RammerSmith} J.~Rammer and H.~Smith, 
                      Rev.\ Mod.\ Phys. {\bf 58}, 323 (1986).

\end{thebibliography}
\end{document}